\documentclass[aps,%jmp, %letter,
scriptaddress,spanish,%prl
] %,twocolumn]
{revtex4}  %% twocolumn  %% para dos columnas
\usepackage{amsmath, amssymb}
\usepackage{makeidx}
\usepackage{amsfonts}
\usepackage{ucs}
\usepackage[utf8]{inputenc}
\usepackage[usenames,dvipsnames]{pstricks}
\usepackage{subfigure}  % use for side-by-side figures
\usepackage{epsfig}
\usepackage{pst-grad} % For gradients
\usepackage{pst-plot} % For axes
\usepackage[colorlinks,hyperindex]{hyperref}
\usepackage{multirow}   %%% for tables multirow
\hypersetup
{
	colorlinks,%
	citecolor=black,%
	linkcolor=black,%
	urlcolor=black,%
}

\newcommand{\ms}{\medskip}
\newcommand{\tcr}{\textcolor{red}}
\newcommand{\tcb}{\textcolor{blue}}

\makeindex

\begin{document}

	\title{\large Superfluid Rayleigh-Plesset extension of FLRW cosmology}
%Cosmological solutions in FLRW cosmology modified with a pair of cubic and sixth powers of the inverse of $a$} %$1/a$}
	
	\author{Haret C. Rosu$^*$,  Stefan C. Mancas$^{\diamond}$, Chun-Chung Hsieh$^\star$}
	\affiliation{$^*$~IPICYT- Instituto Potosino de Investigación Científica y Tecnológica, Camino a la Presa San José 2055, Col. Lomas 4a Sección, San Luis Potosí, 78216 S.L.P., Mexico\\
$^\diamond$~Department of Mathematics, Embry-Riddle Aeronautical University, Daytona Beach, FL 32114-3900, USA\\
$^\star$~Institute of Mathematics, Academia Sinica, Nankang, Taipei 115, Taiwan}
   % \affiliation{$^\diamond$~Math Dept, Embry-Riddle Aeronautical University, Daytona Beach, FL, USA}

	\date{1 May 2021}
	
	\begin{abstract}

Guided by the analogy with the Rayleigh-Plesset dynamics of multielectron bubbles in superfluid He-4,
we consider the cosmological FLRW evolution equation with additional cubic and sixth powers of the inverse of the scale factor of the universe.
For the barotropic parameter $w=2/3$ (coasting universe), along with zero cosmological constant in the absence of viscous terms, by using the Sundman time as evolution parameter, we present parametric solutions for the scale factor of the universe in terms of rational expressions of Weierstrass elliptic functions and their particular cases thereof. For other values of the equation of state parameter $w$, such as $w=-1$, but also the same coasting case,
we present a more standard discussion in the conformal time variable using solutions obtained by numerical integration.%\\ %\hfill
$\qquad$ %\flushright 
\tcr{{\scriptsize Annals of Physics 429 (2021) 168490} $\,\,$ \tcb{{\scriptsize arXiv:2010.01720v4}}}
\end{abstract}

\maketitle

\section{Introduction}
\label{sec:intro}

Hydrodynamic phenomena have long served as rich-in-wisdom analogies for astrophysics and cosmology. The Mach shock wave context of the black hole evaporation introduced by Unruh \cite{u81} four decades ago is just a famous example which attracted a lot of attention, but clearly it is not the only case. Superfluid hydrodynamics with its embodiments as Bose-Einstein condensates and superlight bosons is another notorious example in strong association with galactic halos and the concept of dark matter \cite{sin,zhang,hui},
whereas various other superfluid analogs of quantum field cosmic phenomena can be found in a review by Volovik \cite{vol}. Let us also recall that the idea of a superfluid universe has been central in the works of the late Professor Kerson Huang who summarized the publications with his collaborators along many years in an inspiring book on this topic \cite{Huangbook}.
%%\footnote{The book was ready for the shelves just one month after the passing away of Professor Huang on September 1, 2016.}

\medskip

On the other hand, in the latter case of cosmology, the fluid and classical mechanics analogies start to abound \cite{Bini08,Far20,Far20u,SP21}. Among the fluid analogies, a less known, but equally promising one has been recently introduced by Rousseaux and Mancas \cite{rm20}, that we address here from the superfluid point of view. It is based on the similarities between the Rayleigh-Plesset (RP) equation \cite{Ray1917,Pless1949}, which is a reshaped form of the Navier-Stokes equations describing the dynamical evolution of bubbles in fluids \cite{vanGorder2016}, and the cosmological dynamics as introduced by Friedmann, Lema\^{\i}tre, Robertson, and Walker almost one century ago. The modern compelling components of the universe, such as dark energy and dark matter, may require the addition of supplementary terms to the latter equations. This is what Banerjee et al \cite{b18} have recently done by adding a fourth power term in the inverse of the scale factor as due to dark energy to enhance the late expansion rate of the universe, while Rousseaux and Mancas \cite{rm20} speculated on the possible addition of more viscosity terms in the cosmological evolution, motivated by the case of the RP equation for which the addition of this kind of terms is a common usage, as revealed by some references cited in \cite{rm20}.

\medskip

Extensions of the FLRW cosmology are generically motivated by the dark energy era \cite{Bambaetal}, but in this paper, motivated by \cite{rm20}, we investigate in an explicit way the addition of polynomial nonlinearities to the usual FLRW evolution equation. In particular, we are interested not only in the common cubic power of the usual RP equation, but also in the effects of an additional nonlinear power of order six which occurs in the Rayleigh-Plesset framework developed in the study of multielectron bubbles (henceforce MEBs) in superfluid Helium-4 laboratory physics \cite{sw81,rev2007,rev2020}. Such a term takes into account
the repulsive Coulomb interaction of the electrons forming a thin layer at the surface of the bubbles. By considering such a phenomenological detail in theoretical cosmology, we place ourselves from the beginning in a particular case of the class of inhomogeneous fluid models of the universe which are widely discussed in the literature, see e.g. \cite{nojod05} and references therein. However, we believe that the detail that we emphasize here may be important not only for the laboratory phenomenology, but also in cosmology and astrophysics.

\ms

We also remark that the addition of power nonlinearities is well known in other research areas. It is sufficient to recall that a quintic power is added to the cubic nonlinear Schr\"odinger equation to handle the rich phenomenology of soliton propagation in nonlinear waveguides, and the case of laboratory BECs, where the cubic-quintic Ginzburg-Landau equation has been for several decades the standard equation for modeling the BEC behavior. Differently from the case of the single cubic nonlinearity which is integrable by the inverse scattering method, the solutions of the cubic quintic cases can develop blow-ups, finite time singularities where the amplitude of solution reaches infinity in a finite time \cite{RevAK2001}.
The FLRW cosmology is not far from these areas in the dynamical nonlinear perspective. The standard FLRW equations with physical content characterized by normal (not `super-negative'-dark \cite{NO20,C02}) forms of equation of state are integrable, see e.g. \cite{Far99} and the references to textbooks therein, but adding more power nonlinearities can generate future-in-time singularities.

\ms

The rest of the paper is focused on obtaining some representative cosmological scaling factors of this extended FLRW model for zero cosmological constant which are obtained as solutions of the Weierstrass elliptic equation.
In section \ref{sec2}, we establish the nonlinear differential equation corresponding to this generalized FLRW model.
In the case of the coasting universe, by a change of variable of the Sundman type, this generalized FLRW equation is turned into a Weierstrass elliptic equation whose standard mathematical setup is recalled. In section \ref{sec3}, we present some sets of scale factors of the universe obtained as parametric solutions of this model in the form of rational expressions of Weierstrass elliptic $\wp$ functions including the special and degenerate cases \cite{Steiner2007}. In section \ref{sec4}, we work in the standard conformal time variable in which presumably closed-form results cannot be obtained, and display some numerical results for the vacuum case and the same coasting case. Finally, section \ref{sec5} contains some conclusions and further remarks on the results.

\section{The FLRW equation with superfluid Rayleigh-Plesset terms %equation for multielectron bubbles, the cosmological counterpart,
and the elliptic equation}\label{sec2}

 The standard RP equation for the time-evolving radius, $R(t)$, of a bubble in a fluid reads %%(Eq.~(9) in the Rousseaux-Mancas (RM) draft)
\begin{equation}\label{eq1}
\frac{\ddot{R}}{R}+\frac{3}{2}\left(\frac{\dot{R}}{R}\right)^2+\frac{-\Delta P(t)}{\rho_l}\left(\frac{1}{R}\right)^2
+\frac{2\gamma}{\rho_l}\left(\frac{1}{R}\right)^3+4\nu\left(\frac{\dot{R}}{R^3}\right)=0~,
\end{equation}
to which a sixth power term in $1/R$ should be added in the case of MEBs. The symbols in the coefficients are: $\rho_l$ -- the liquid density, $\nu$ -- the liquid kinematic viscosity, $\gamma$ -- the surface tension, and $\Delta P(t)=P_{in}(t)-P_{out}(t)$ -- the relative pressure drop at the bubble with respect to the pressure in the liquid at infinity, and the dot denotes the time derivative. For this equation without the viscosity parametric solutions in terms of rational expressions of Weierstrass $\wp$ functions have been already obtained by Kudryashov and Sinelshchikov \cite{ks15}. %see also \cite{MR2016}.

By comparison, Rousseaux and Mancas \cite{rm20} have proposed a $\chi$ generalization of the single FLRW equation obtained from the coupled FLRW ones
\begin{equation} \label{eq2}
\frac{\ddot{a}}{a}+\chi\left(\frac{\dot{a}}{a}\right)^2
+\chi k c^2\left(\frac{1}{a}\right)^2-\frac{1+\chi}{3}\Lambda c^2
+\alpha\left(\frac{1}{a}\right)^3+\beta\left(\frac{1}{a}\right)^4
+\delta\left(\frac{\dot{a}}{a^3}\right)=0~.
\end{equation}
where $\chi=(1+3w)/2$, $w$ is the parameter entering the barotropic equation of state $p=w\rho c^2$, $\rho$ is the total energy density with the standard vacuum, radiation, and matter components, $k$ is the spatial curvature index, and $\Lambda$ is the cosmological constant. The $\alpha$
and $\delta$ coefficients are cosmological equivalents of the corresponding Rayleigh-Plesset coefficients,
and $\beta$ is the coefficient of the term generated by the scenario developed in \cite{b18}. The equivalence $a\equiv R$ serves as the basis of the analogy.

\medskip

In this article, we study the simplified case when the quartic term is not taken into account, and also the viscosity term is discarded
since in the superfluid regime the
viscosity is negligible. In addition, we will also assume $\Lambda=0$, but on the other hand, we will include the sixth-power MEB term
equipped with a negative coupling parameter $-\zeta$. With these assumptions \eqref{eq2} reads
\begin{equation}\label{eq3}
\frac{\ddot{a}}{a}+\chi\left(\frac{\dot{a}}{a}\right)^2 %+\frac{-\Delta P(t)}{\rho_l}
+\chi kc^2\left(\frac{1}{a}\right)^2 %-\frac{1+\Omega}{3}\Lambda c^2 %+\frac{2\gamma}{\rho_l}
+\alpha\left(\frac{1}{a}\right)^3 %+4\nu
-\zeta\left(\frac{1}{a}\right)^6=0~.
\end{equation}
We consider the case $\chi=3/2$ ($w=2/3$) which is in direct correspondence with the RP equation and represents the case of monoatomic gas. Using natural units for the speed of light, $c=1$, we write \eqref{eq3} as
\begin{equation} \label{eq4}
a \ddot  a +\frac{3}{2}{\dot a}^2=\zeta a^{-4}-\alpha a^{-1}-\frac 3 2 k~.
\end{equation}
Multiplying \eqref{eq4} by the integrating factor $2a^2 \dot a$ and integrating once,  we obtain
\begin{equation} \label{eq5}
{\dot a}^2=-2 \zeta a^{-4}+ 4c_1a^{-3}-\alpha a^{-1}-k,
\end{equation}
where $c_1$ is an arbitrary constant of integration
that is found from two initial conditions.
For $\dot{a}(0)=0$ and ${a(0)}=a_0$ then
$c_1=c_1(\alpha,k,\zeta)\equiv(2 \zeta+\alpha {a_0}^3+k {a_0}^4)/4 a_0$. %}
Notice that $a_0$ is a root of the right hand side of (\ref{eq5}), and also of the polynomial $Q(a)$ in (\ref{eq6}) below.
We shall use the symbol $\mu$ for $a_0$ in the rest of the paper. Most of the solutions that will be displayed in this paper do not have the big bang
initial condition $a_0=\mu=0$, but the initial Hubble parameter $H(0)=\dot{a}(0)/\mu$ is zero.
%\textcolor{red}{and also one can tune $c_1$ to get an appropriate $\mu$.}\textcolor{blue}{$c_1$ is obtained from the initial conditions and not vice-versa}

Using the Sundman transformation $dt=a^2 d\tau$ we obtain the elliptic equation
\begin{equation} \label{eq6}
{a_\tau}^2=-k a^4 -\alpha a^3 +4 c_1 a-2 \zeta\equiv Q(a)~,
\end{equation}
where the $\tau$ subindex denotes the $\tau$ derivative.
This equation can be viewed as an energy conservation equation for a one degree of freedom classical particle of mass $m=2$ if written in the form
\begin{equation}
{a_\tau}^2+V(a)=E,
\end{equation} where the potential energy is
\begin{equation}
V(a)=a(k a^3 +\alpha a^2 -4 c_1)~,
\end{equation}
 and the total energy is
 \begin{equation}
 E(a, \dot a)=-2 \zeta=a^4\dot a^2-4 c_1 a +\alpha a^3+k a^4~.
 \end{equation}
% \textcolor{blue}{
 It is worth noting that the big bang condition $a(0)=0$, implies that the total energy is zero for $\zeta=0$, see Figs.~3(a), 4(a) and 9(c).
\subsection{The General Mathematical Setup of Elliptic Equations}
It is well known \cite{Wei, Whi, AS} that the solutions $a(\tau)$ of
\begin{equation} \label{eq7}
{a_\tau}^2=\mathrm{c}_4 a^4+4 \mathrm{c}_3 a^3+ 6 \mathrm{c}_2 a^2+ 4 \mathrm{c}_1 a+\mathrm{c}_0,
\end{equation}
can be expressed in terms of Weierstrass elliptic functions $\wp(\tau;g_2,g_3)$, which is a solution to
\begin{equation}\label{eq16}
{\wp_\tau}^2=4 \wp^3-g_2 \wp -g _3,
\end{equation}
via the transformation
\begin{equation}\label{eq8}
a(\tau)=\hat a+\frac{ \sqrt{Q(\hat a)}\wp'(\tau-\tau_0;g_2,g_3)+\frac 1 2 {Q}'(\hat a)\Big[\wp(\tau-\tau_0;g_2,g_3)-\frac {1}{24}{Q}''(\hat a)\Big]+\frac{1}{24}Q(\hat a){Q}^{(3)}(\hat a)}{2\Big[\wp(\tau-\tau_0;g_2,g_3)-\frac{1}{24}{Q}''(\hat a)\Big]^2-\frac{1}{48}Q(\hat a)Q^{(4)}(\hat a)},
\end{equation}
where $\hat a$ is not necessarily a root of $Q(a)$, and $g_2,g_3$ are elliptic invariants of $\wp(\tau)$, which satisfy
\begin{equation}\label{eq9}
\begin{aligned}
	g_2&=\mathrm{c}_4 \mathrm{c}_0 -4 \mathrm{c}_3 \mathrm{c}_1 +3 {\mathrm{c}_2}^2=2 k \zeta+c_1\alpha\\
	g_3 &=\mathrm{c}_4 \mathrm{c}_2 \mathrm{c}_0 +2 \mathrm{c}_3 \mathrm{c}_2 \mathrm{c}_1 -\mathrm{c}_4 {\mathrm{c}_1}^2-{\mathrm{c}_2}^3-{\mathrm{c}_3}^2\mathrm{c}_0=k {c_1}^2+\frac 1 8 \alpha^2\zeta.\\
\end{aligned}
\end{equation}
These invariants, with  the varying parameter $c_1(\alpha,k,\zeta)$ together with the modular discriminant
\begin{equation}\label{eqmod}
\Xi={g_2}^3-27 {g_3}^2
\end{equation}
are used to classify the solutions of  \eqref{eq6}.
In particular, choosing $\mu$ to be any real  root of $Q(a)$, i.e. $Q(\mu)=0$, then the general solution \eqref{eq8} takes the much simpler form
%............................
\begin{equation}  \label{eq10}
a(\tau)=\mu+\frac{Q_a(\mu)}{4 \wp (\tau-\tau_0; g_2;g_3)-\frac{Q_{aa}(\mu)}{6} }~.
\end{equation}
This solution can be also interpreted as the elliptic modulation, $M_Q(\wp;c_1,\mu)$, of any solution of the polynomial $Q(a)$ if we write it in the form
$$
a(\tau)=\mu\left(1+\frac{\frac{Q_a(\mu)}{\mu}}{4 \wp (\tau-\tau_0; g_2;g_3)-\frac{Q_{aa}(\mu)}{6} }\right)\equiv \mu\,M_Q(\wp;c_1,\mu)~,
$$
and can be generalized using (\ref{eq8}) to any point $\hat{a},\hat{\tau}$ of the plane ($a,\tau$).

%%%%%%%%%%%%%%%%%%%%%%%%%%%%%%%%%%%%%%%%%%%%%%%%%%%%%%%%%%%%%%%%%%%%%%%%%%%%%%%%%%%%%%%%%%%%%%%%%%%%% CURVED SPACE
\section{The parametric solutions}\label{sec3}
We are now ready to present a whole wealth of parametric solutions of this extended FLRW model using $\tau$ as parameter.
Although all possible cases can be enlisted in a systematic manner, for a superfluid universe or one containing superfluid bubbles one should consider the simultaneous presence of the Laplace pressure and electrostatic terms, i.e., both $\alpha$ and $\zeta$ should not be zero simultaneously. When both of them are chosen to be zero it corresponds to the standard FLRW cosmology. The case when one of them is zero such as $\alpha=0$ and $\zeta\neq 0$ are not physical, while $\alpha\neq 0$ and $\zeta=0$ describes common liquids, and thus will not be of interest here.

\subsection{Curved Space Cases, $k\neq 0$, $\Xi \ne 0$}
Substituting  $Q(\mu)$ and its derivatives,
(\ref{eq10}) becomes the parametric {\it rational Weierstrass}
solution
\begin{equation}\label{eq11}
a(\tau)=\mu+\frac{4c_1-4 k  \mu^3-3\alpha \mu ^2}{4\wp \left(\tau-\tau_0; 2 k \zeta+\alpha c_1 ;k {c_1}^2+\frac 1 8 \alpha^2\zeta\right)+2 k{\mu}^2+\alpha \mu}~,   \quad \alpha \ne 0, \quad \zeta \ne 0~.
\end{equation}

Next, we choose two particular cases of nondegenerate solutions, one referred as superfluid rational lemniscatic for $g_3=0$, and the other as standard rational equianharmonic for $g_2=0$ because it corresponds to the standard FLRW cosmology for which $\alpha=\zeta=0$.

{\bf i)} The lemniscatic case:
For $c_1=\pm \frac{\alpha}{4}\sqrt{-\frac{2\zeta}{k}}$ %}
obtained by solving $g_3=0$, and $\mu$ a root of   %\tcb{
$-k a^4  -\alpha a ^3 \pm \alpha \sqrt{-\frac{2 \zeta}{k}}a-2 \zeta =0$ %}
we obtain the rational {\it lemniscatic} solution
\begin{equation}\label{eq17}
a(\tau)=\mu+\frac{\pm \alpha\sqrt{-\frac{2\zeta}{k}}-4 k  \mu^3-3\alpha \mu ^2}{4\wp \left(\tau-\tau_0; 2 k \zeta \pm \frac{\alpha^2}{4}\sqrt{-\frac{2\zeta}{k}}; 0\right)+2 k{\mu}^2+\alpha \mu}~,
\quad \alpha \ne 0, \quad \zeta \ne 0, \quad g_3=0.
\end{equation}

\begin{figure}[h!]
\centering
\subfigure[\ $a(\tau)=\mu+\frac{\sqrt{2}-4\mu^3-3\mu^2}{4\wp(\tau;-2+\frac{1}{2\sqrt{2}};0)+2\mu^2+\mu}$ for $\mu=1.189$
and $-a(\tau)$ for $\mu=-1.189$ and sign flips $(+,+)\rightarrow (-,-)$ in $(\alpha,c_1)$.]{
\includegraphics[scale=0.575]{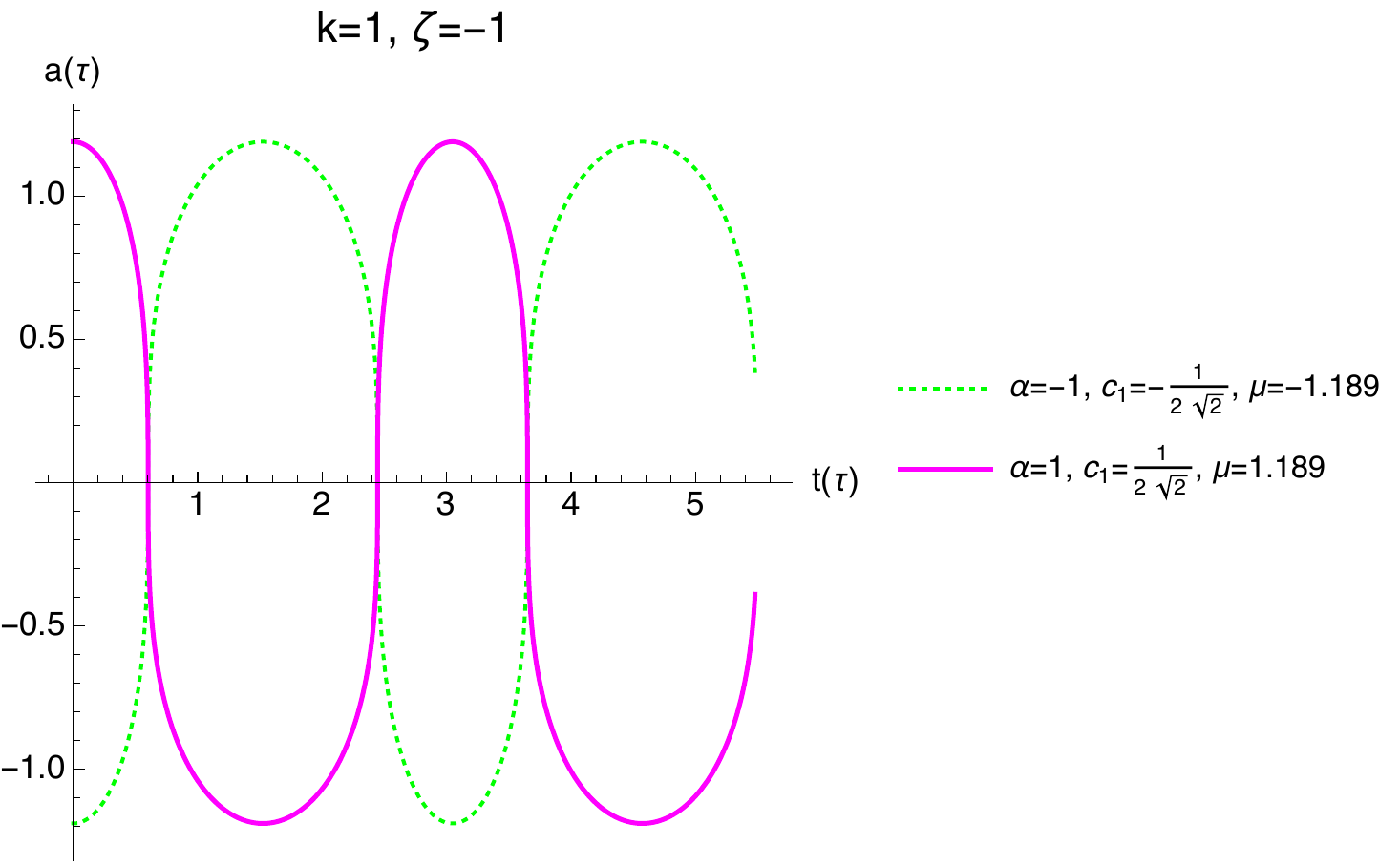}}
\subfigure[\ $a(\tau)=\mu+\frac{-\sqrt{2}-4\mu^3+3\mu^2}{4\wp(\tau;-2-\frac{1}{2\sqrt{2}};0)+2\mu^2-\mu}$ for $\mu=1.189$
and $-a(\tau)$ for $\mu=-1.189$ and sign flips $(-,-)\rightarrow (+,+)$ in $(\alpha,c_1)$..]{
\includegraphics[scale=0.575]{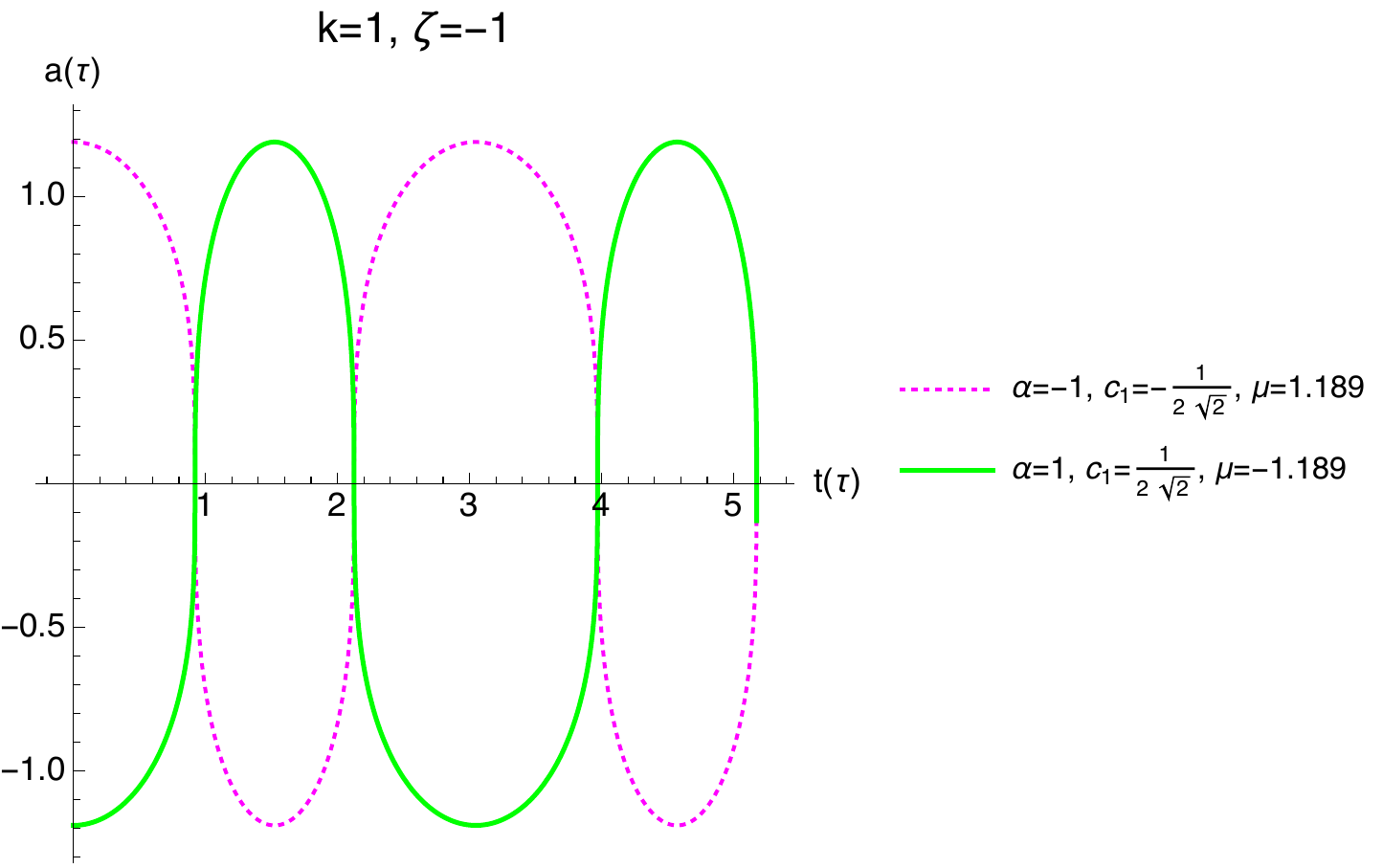}}
\subfigure[\ $a(\tau)=\mu+\frac{-\sqrt{2}-4\mu^3+3\mu^2}{4\wp(\tau;-2-\frac{1}{2\sqrt{2}};0)+2\mu^2-\mu}$
for $\mu=1.790$ and $-a(\tau)$ obtained for $\mu=-1.790$ and sign flips $(-,+)\rightarrow (+,-)$ in $(\alpha,c_1)$.]{
\includegraphics[scale=0.575]{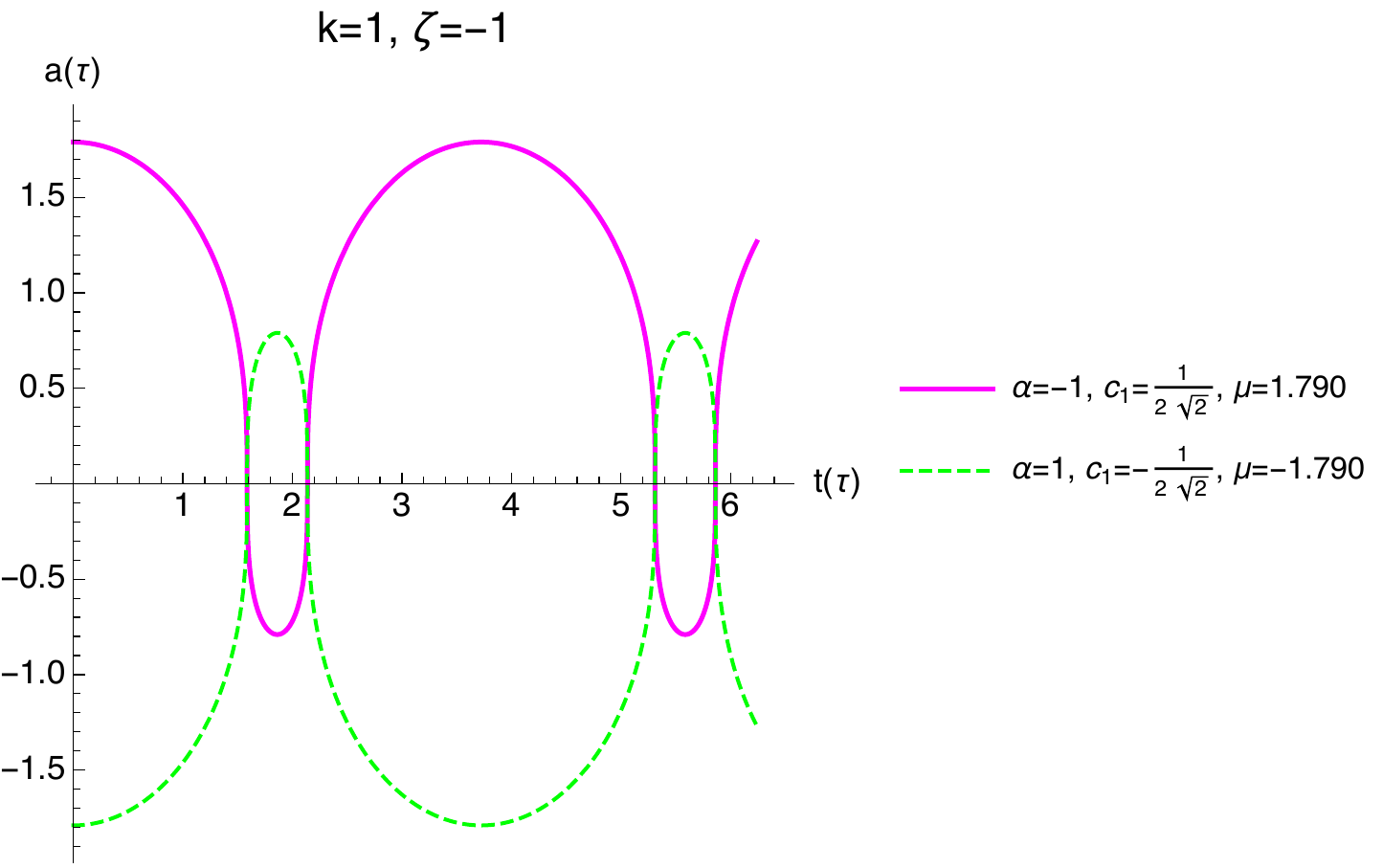}}
\subfigure[\ $a(\tau)=\mu+\frac{-\sqrt{2}-4\mu^3-3\mu^2}{4\wp(\tau;-2-\frac{1}{2\sqrt{2}};0)+2\mu^2+\mu}$
for $\mu=0.790$ and $-a(\tau)$ obtained for $\mu=-0.790$ and sign flips $(+,-)\rightarrow (-,+)$ in $(\alpha,c_1)$.]{
\includegraphics[scale=0.575]{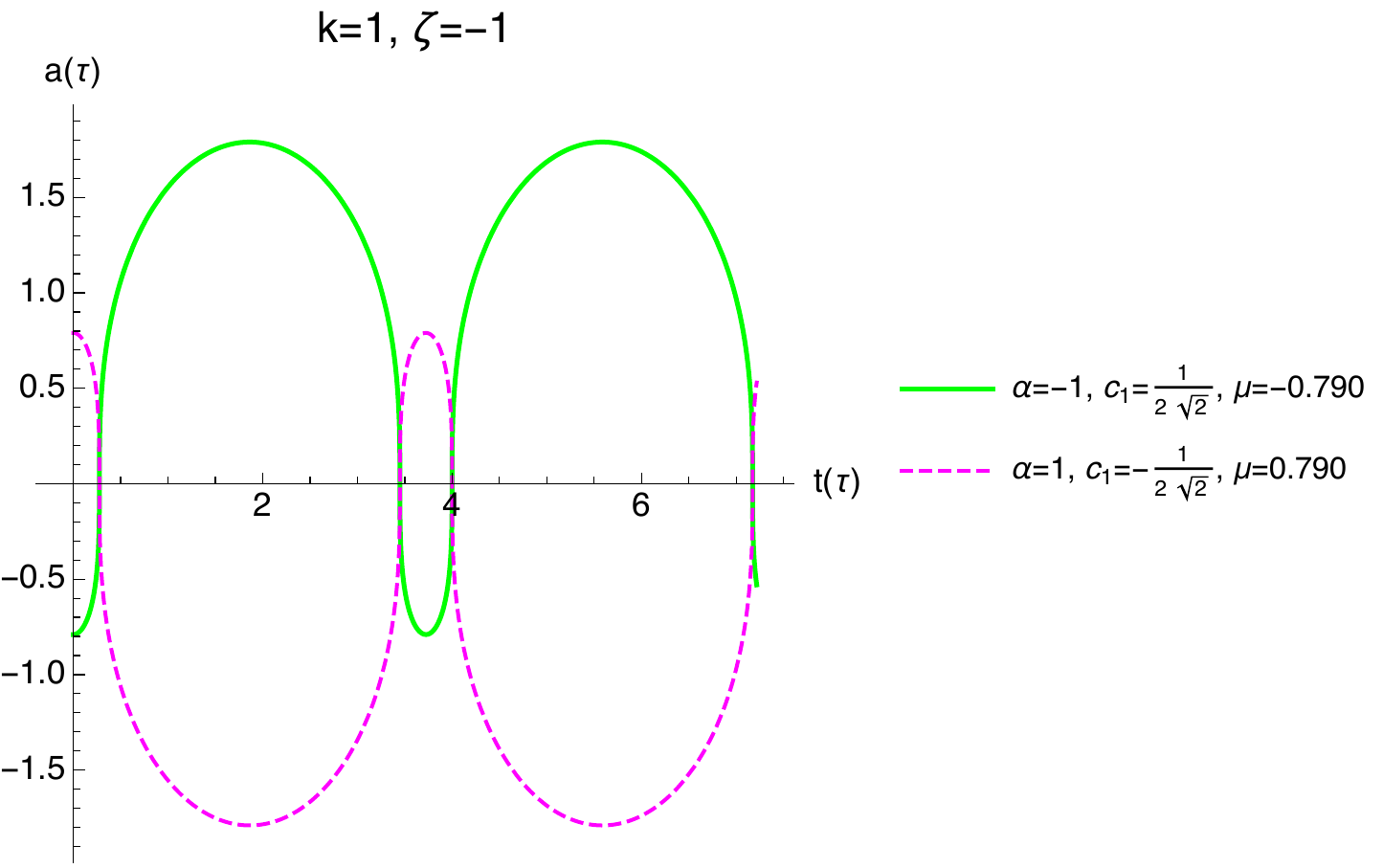}}
\caption{\label{Figura17ab} Rational lemniscatic solutions for a superfluid FLRW closed universe from (\ref{eq17}).}
\end{figure}

\begin{figure}[h!]
\centering
\subfigure[\ $a(\tau)=\mu+\frac{\mp\sqrt{2}+4\mu^3\mp3\mu^2}{4\wp(\tau;-2-\frac{1}{2\sqrt{2}};0)-2\mu^2\pm\mu}$
%%for $\mu=1.790$ and the negative mirror curve $a(\tau)=\mu+\frac{\sqrt{2}+4\mu^3+3\mu^2}{4\wp(\tau;-2-\frac{1}{2\sqrt{2}};0)-2\mu^2-\mu}$
for $\mu=\pm1.790$, respectively.]{
\includegraphics[scale=0.55]{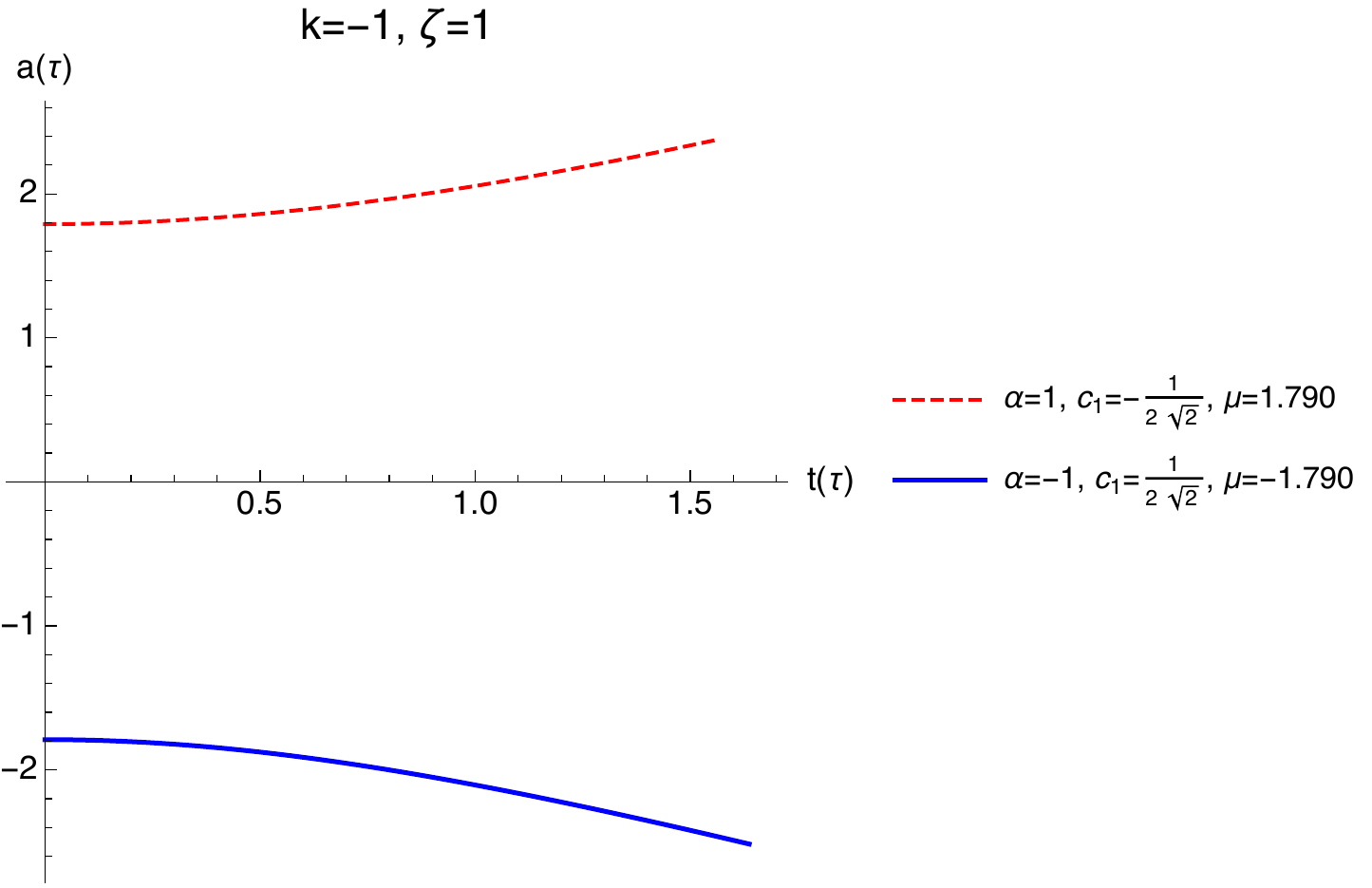}}
\subfigure[\ $a(\tau)=\mu+\frac{\mp\sqrt{2}+4\mu^3\mp3\mu^2}{4\wp(\tau;-2-\frac{1}{2\sqrt{2}};0)-2\mu^2\pm\mu}$ for $\mu=\pm 0.790$, respectively.]{
\includegraphics[scale=0.55]{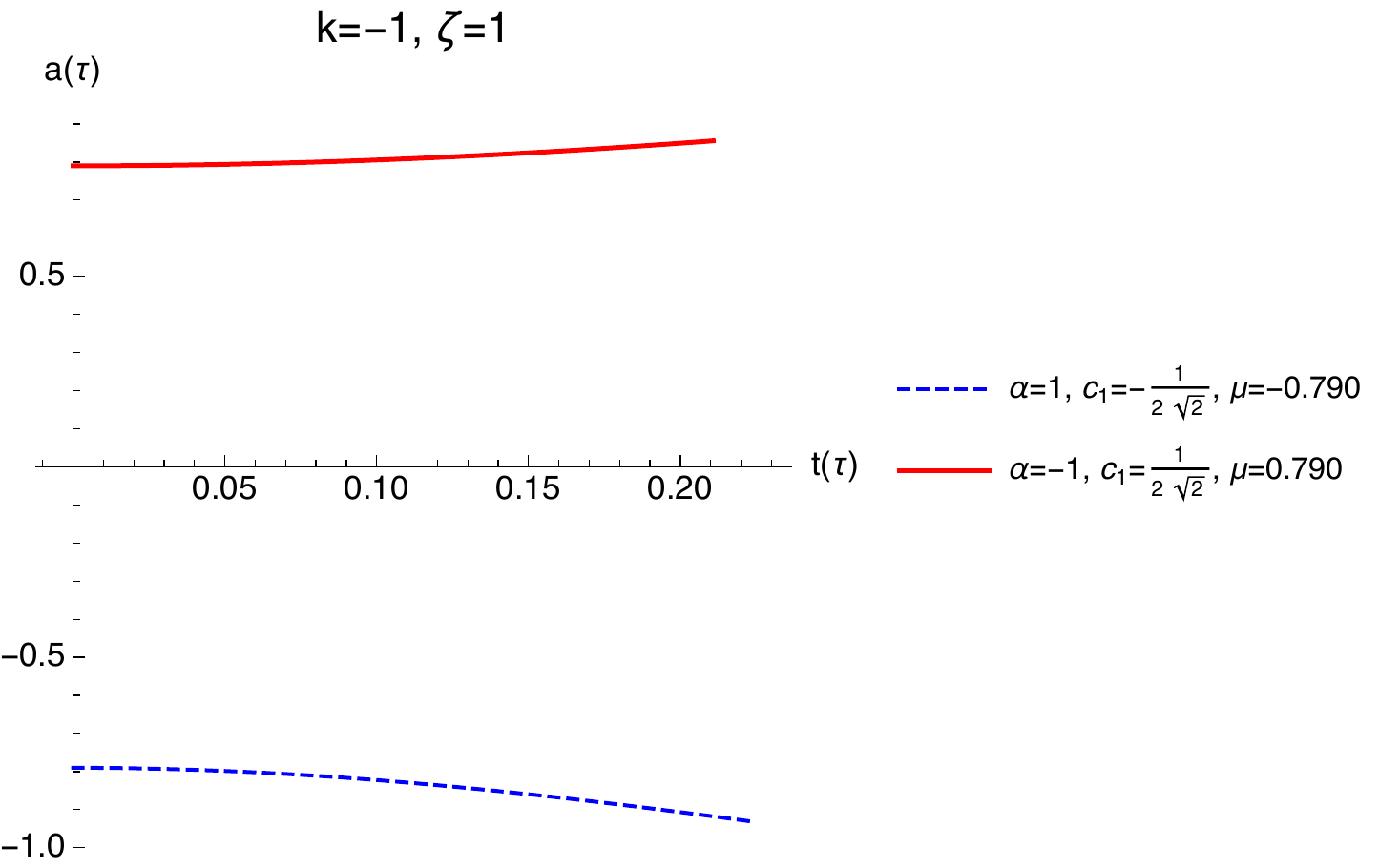}}
\subfigure[\ $a(\tau)=\mu+\frac{\sqrt{2}+4\mu^3-3\mu^2}{4\wp(\tau;-2+\frac{1}{2\sqrt{2}};0)-2\mu^2+\mu}$ for $\mu=1.189$ and
$-a(\tau)$ for $\mu=-1.189$ and sign flips $(+,+)\rightarrow (-,-)$ in $(\alpha,c_1)$.]{
\includegraphics[scale=0.55]{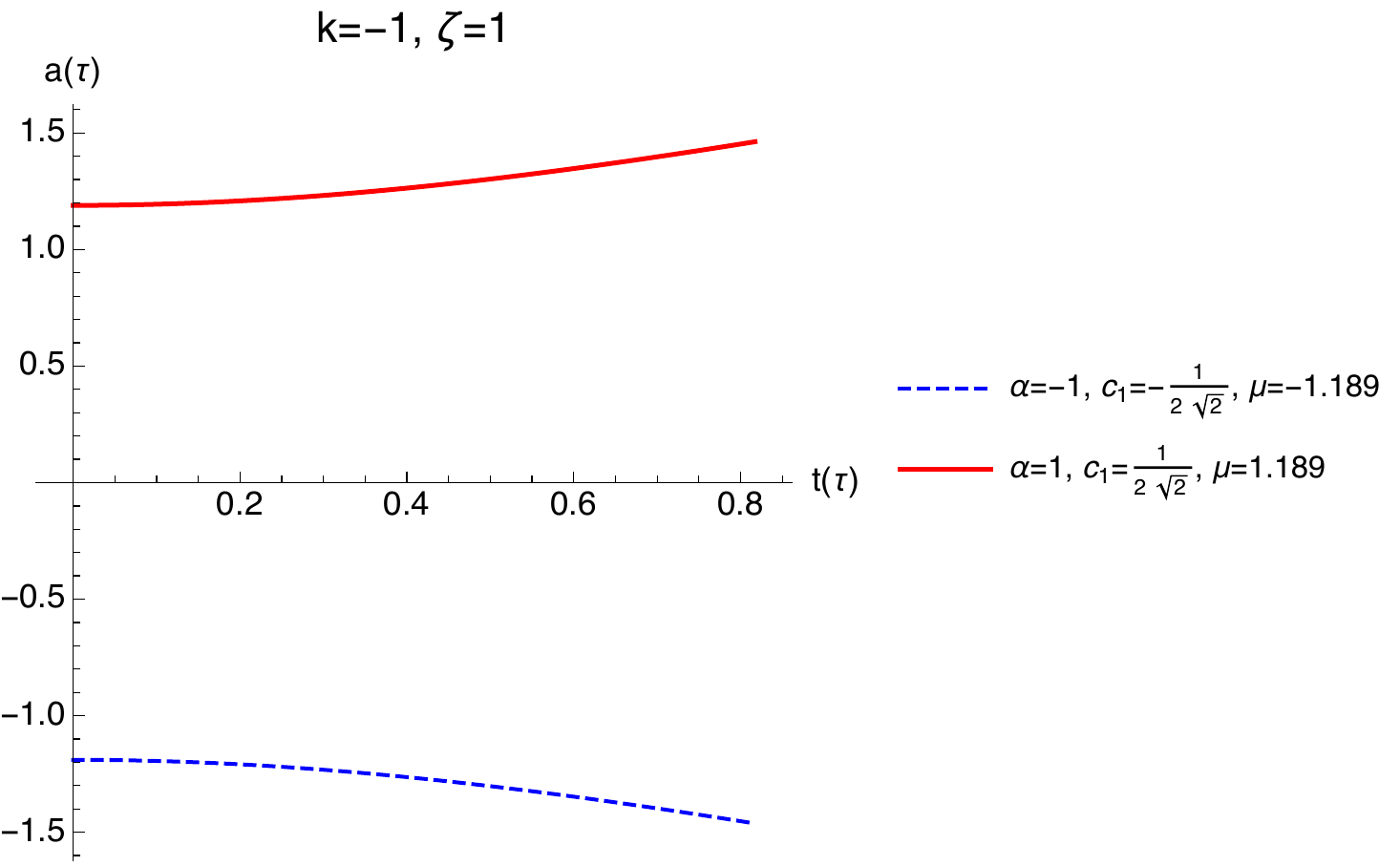}}
\subfigure[\ $a(\tau)=\mu+\frac{-\sqrt{2}+4\mu^3+3\mu^2}{4\wp(\tau;-2+\frac{1}{2\sqrt{2}};0)-2\mu^2-\mu}$ for $\mu=1.189$ and
$-a(\tau)$ for $\mu=-1.189$ and sign flips $(-,-)\rightarrow (+,+)$ in $(\alpha,c_1)$.]{
\includegraphics[scale=0.55]{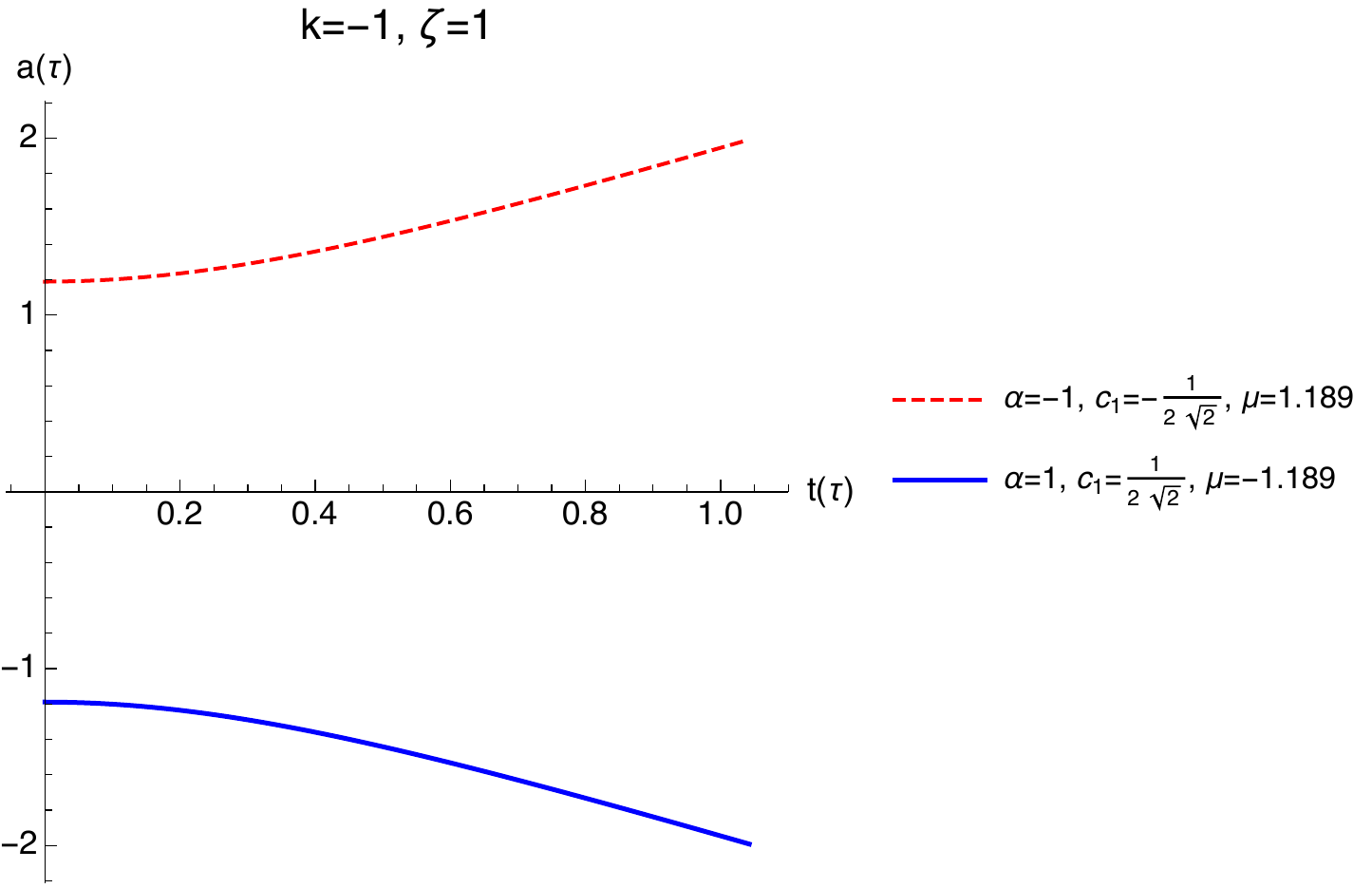}}
\caption{\label{Figura17cd} The same type of solutions for the superfluid open universe from (\ref{eq17}).}
\end{figure}

Plots of the periodic (closed universe) and non periodic (open universe) lemniscatic cases are displayed in Figs.~\ref{Figura17ab} and \ref{Figura17cd}, respectively. One can see the very interesting feature that the physical solutions of the periodic cases have contributions from the initially negative solutions.

\medskip

%%%%%%%%SUBFIGURE 17-ab %%%%%%%%%%%%%%

%%%%%%%%%SUBFIGURE 17-cd %%%%%%%%%%%%%%

\medskip

{\bf ii)}
Standard FLRW cosmology,
$\alpha=0$ and  $\zeta=0$:   Let $\mu$ be  a root of $-k a^4  +4 c_1 a=0$, then \eqref{eq11}
becomes the {\it equianharmonic} solution
\begin{equation}\label{eq16bis}
a(\tau)=\mu +\frac{2(c_1-k \mu^3)}{2\wp \left(\tau-\tau_0; 0;k {c_1}^2\right)+k \mu^2}~,
\quad \alpha = 0, \quad \zeta = 0, \quad g_2=0.
\end{equation}

\begin{figure}[h!]
\centering
\subfigure[\ $a(\tau)=\pm 1/\wp(\tau;0;1)$ for $\mu=0$.]{
\includegraphics[scale=0.575]{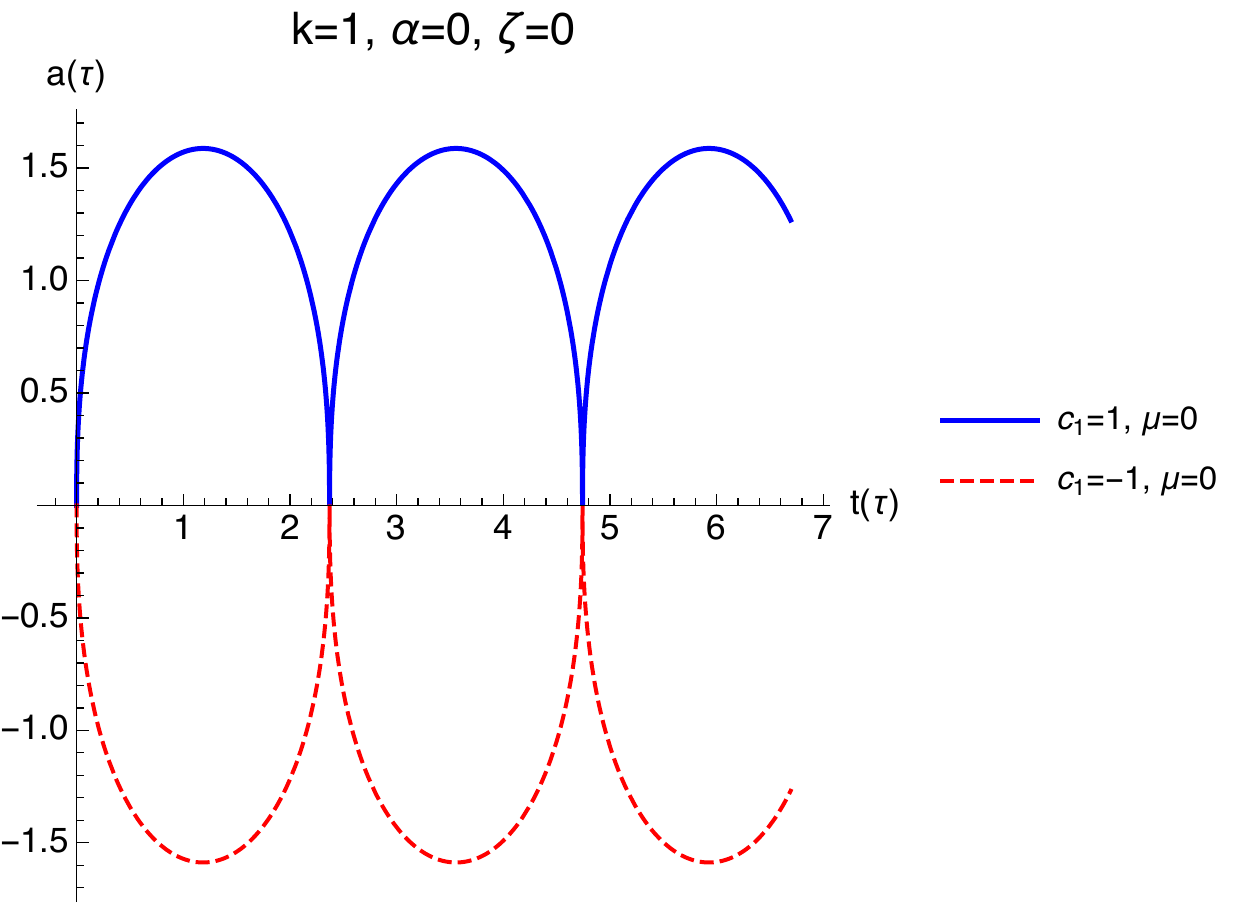}}
\subfigure[\ $a(\tau)=\pm \left(\mu+1/\wp(\tau;0;1)\right)$, for $\mu=\pm 1.587$, respectively.]{
\includegraphics[scale=0.575]{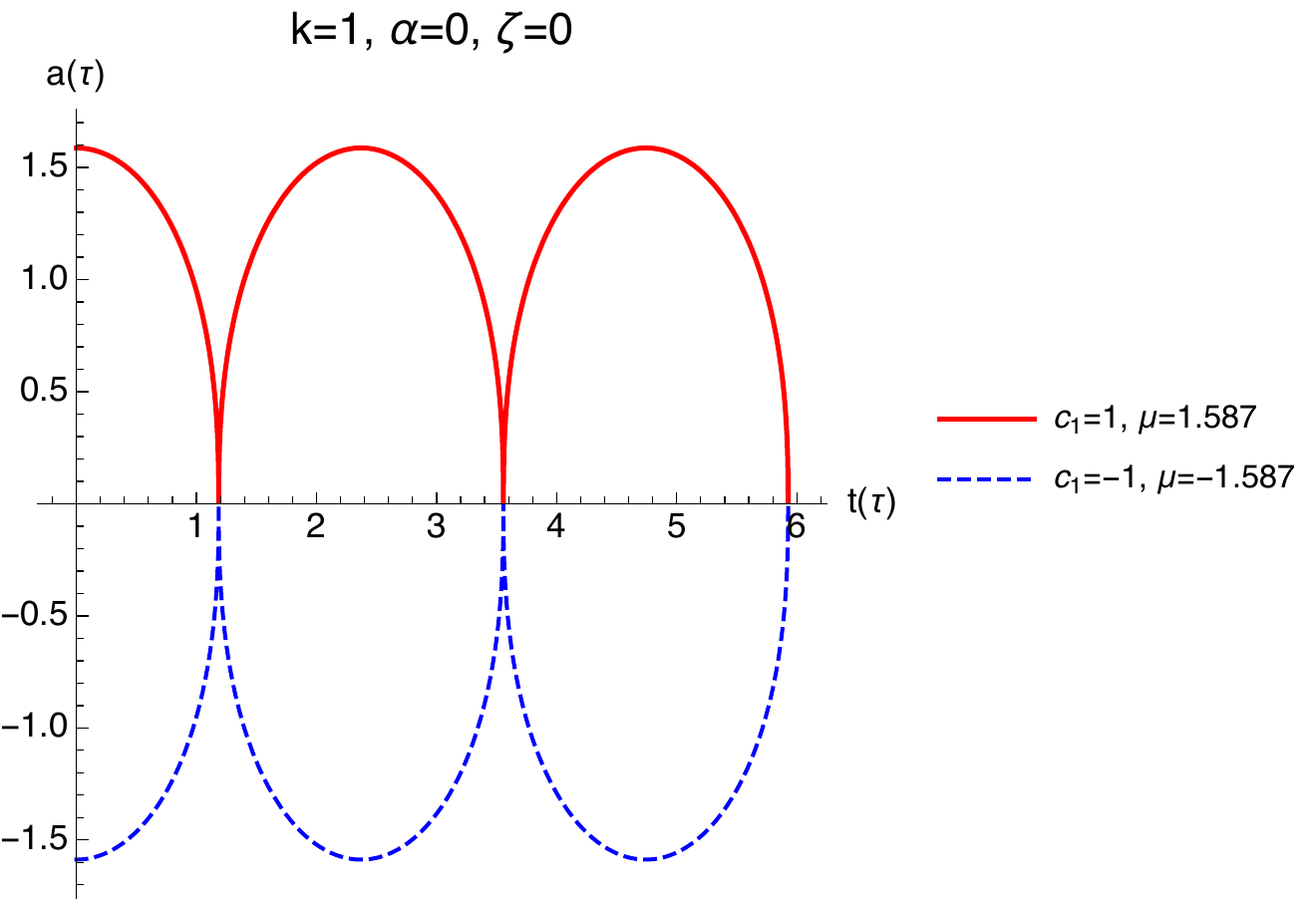}}
\caption{\label{Figura16a} Rational equianharmonic solutions for a standard FLRW ($\alpha=0,\,\zeta=0$) closed universe from (\ref{eq16bis}).}
\end{figure}

\begin{figure}[h!]
\centering
\subfigure[\ $a(\tau)=\pm 1/\wp(\tau;0;-1)$ for $\mu=0$ by sign flip of $c_1$.]{
\includegraphics[scale=0.575]{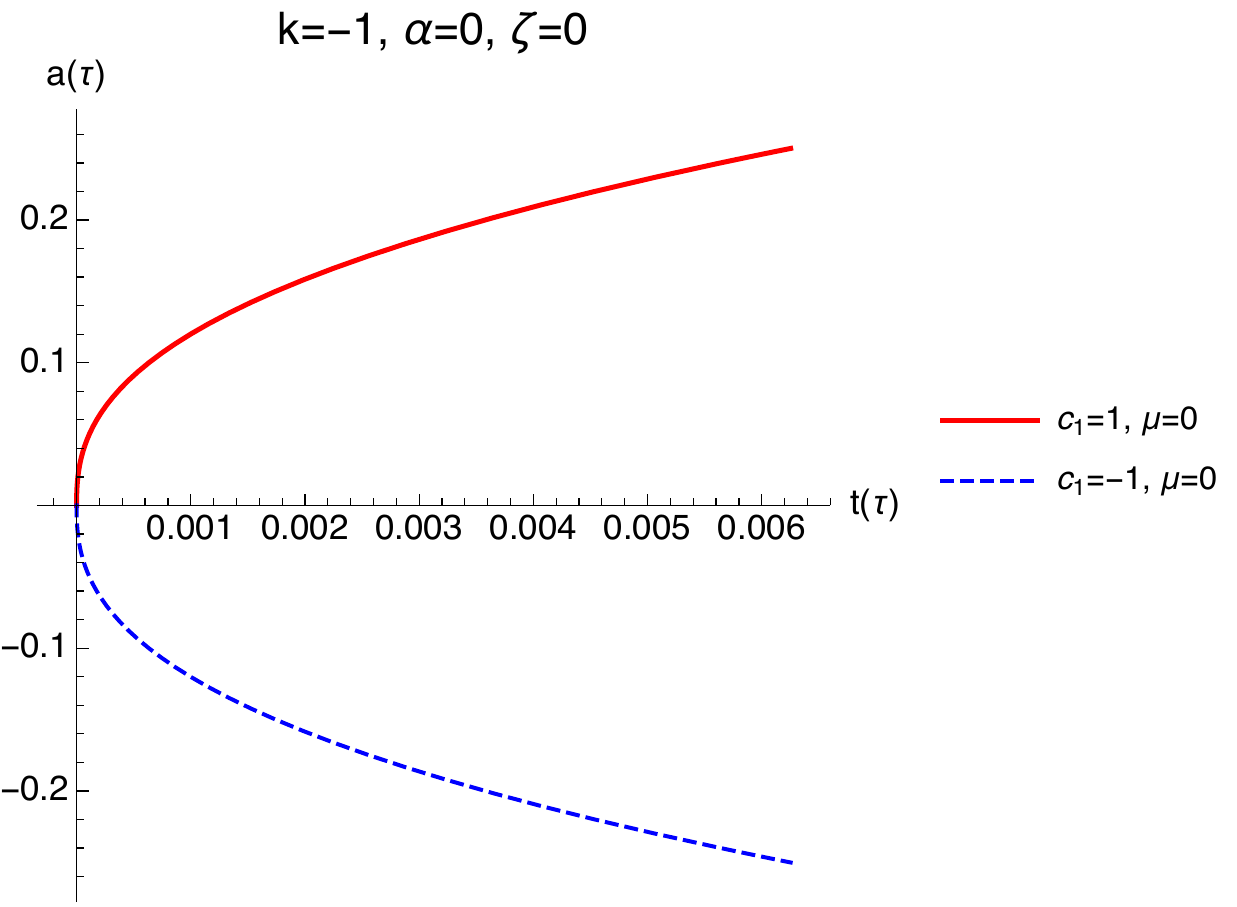}}
\subfigure[\ $a(\tau)=\mu+\frac{2(1+\mu^3)}{2\wp(\tau;0;-1)-\mu^2}$ for $\mu=1.587$ and $-a(\tau)$ for $\mu=-1.587$ and sign flip $+ \rightarrow -$ of $c_1$, respectively.]{
\includegraphics[scale=0.575]{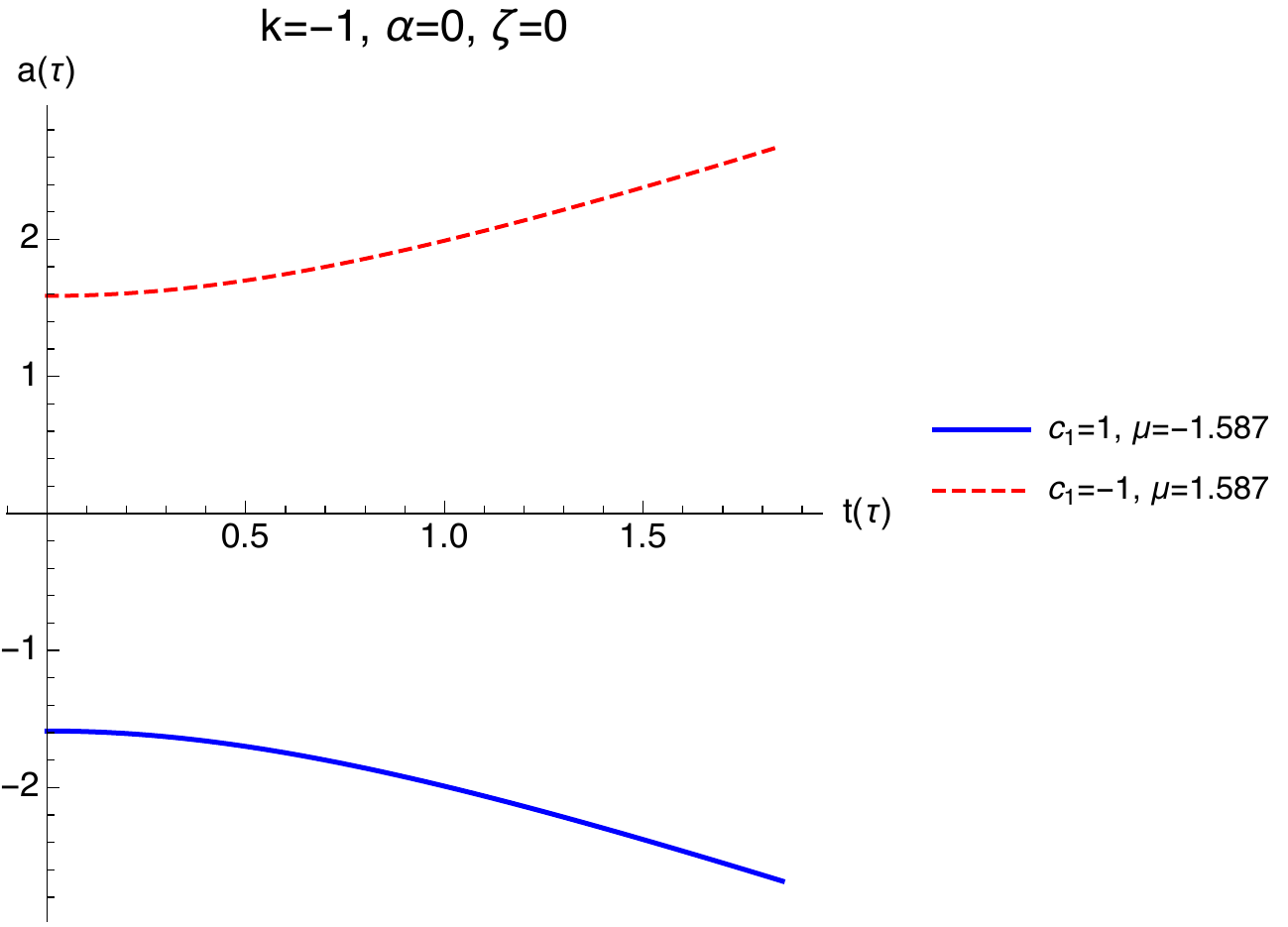}}
\caption{\label{Figura16b}
Same type of solutions for a standard FLRW ($\alpha=0,\,\zeta=0$) open universe from (\ref{eq16bis}).}
\end{figure}

The equianharmonic periodic case corresponding to the closed universe is plotted in Fig.~\ref{Figura16a}, while the non periodic case of the open universe is displayed in Fig.~\ref{Figura16b}. As can be seen in these figures, the solutions are in the two classes of either pure positive (physical) or pure negative (nonphysical).

\medskip

%%%%%%%%SUBFIGURE 16-a %%%%%%%%%%%%%% FIG 3

%%%%%%%%%SUBFIGURE 16-b %%%%%%%%%%%%%% FIG 4

\subsection{Degenerate Modular Cases, $k\neq 0$, $\Xi=0$}

We now study the degenerate cases when $\Xi=0$  by finding numerically the coefficient $c_1$ by solving $\Xi=0$.
In this case  $\wp(\zeta,g_2,g_3)$ degenerates into {\it hyperbolic},  {\it trigonometric}  functions, or {\it rational} functions, and \eqref{eq11} can be simplified further as follows:

{\bf i)}   Let $g_2=12{\hat e}^2>0$ and $g_3=-8{\hat e}^3<0$;
 the Weierstrass $\wp$ function is  simplified to $\wp(\tau;12{\hat e}^2,-8{\hat e}^3)=\hat e\left[1+3~ \mathrm{csch}^2(\sqrt{3\hat e}~\tau)\right]$.
Using this in  \eqref{eq11}  we obtain  the  rational {\it hyperbolic} solutions
\begin{equation}  \label{eq19}
a(\tau)=\mu+\frac{4c_1^*-4 k  \mu^3-3\alpha \mu ^2}{ 2\sqrt[3] {-g_3}\left[1+3\mathrm{csch}^2\left(\frac{\sqrt 6}{2}\sqrt[6]{-g_3}(\tau-\tau_0)\right)\right]+2 k{\mu}^2+\alpha \mu}, %\quad k \ne 0, \quad \Xi = 0,
\quad g_2>0, \quad g_3<0.
\end{equation}

\begin{figure}[h!]
\centering
\subfigure[\ $a(\tau)=\mu +\frac{7-8\mu^3-15\mu^2}{8\left(1+3\mathrm{csch}^2\left(\frac{\sqrt{6}}{4}\tau\right)\right)+4\mu^2+5\mu}$
for $\mu=\frac{1}{4}\left(-1\pm\sqrt{33}\right)$.]{
\includegraphics[scale=0.575]{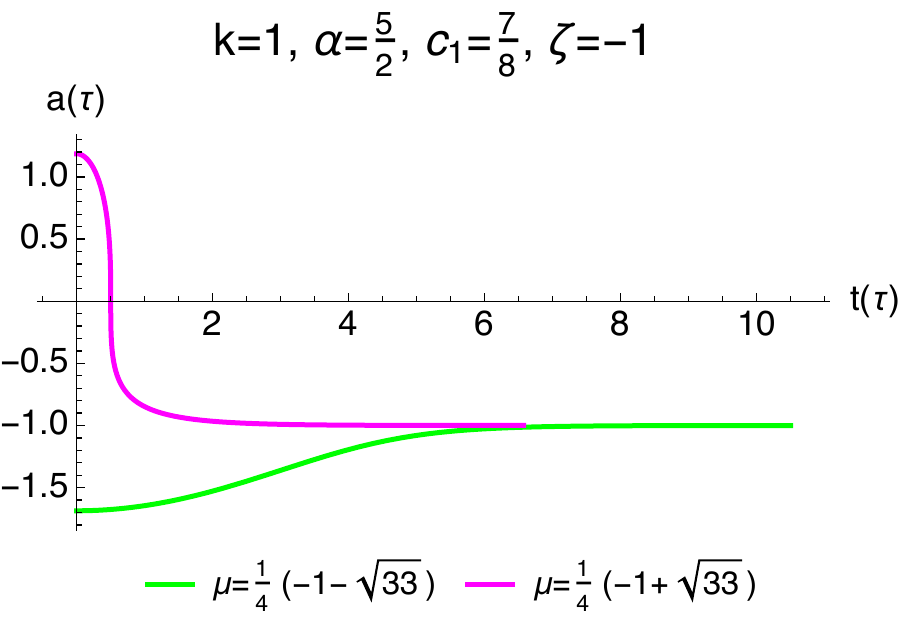}}
\subfigure[\
$a(\tau)=\mu +\frac{\pm 6.116+4\mu^3-9.813\mu^2}{2\left(1+3\mathrm{csch}^2\left(\frac{\sqrt{6}}{2}\,\tau\right)\right)-2\mu^2\pm 3.271\mu}$ for $\mu=\pm 0.347$.]{
\includegraphics[scale=0.575]{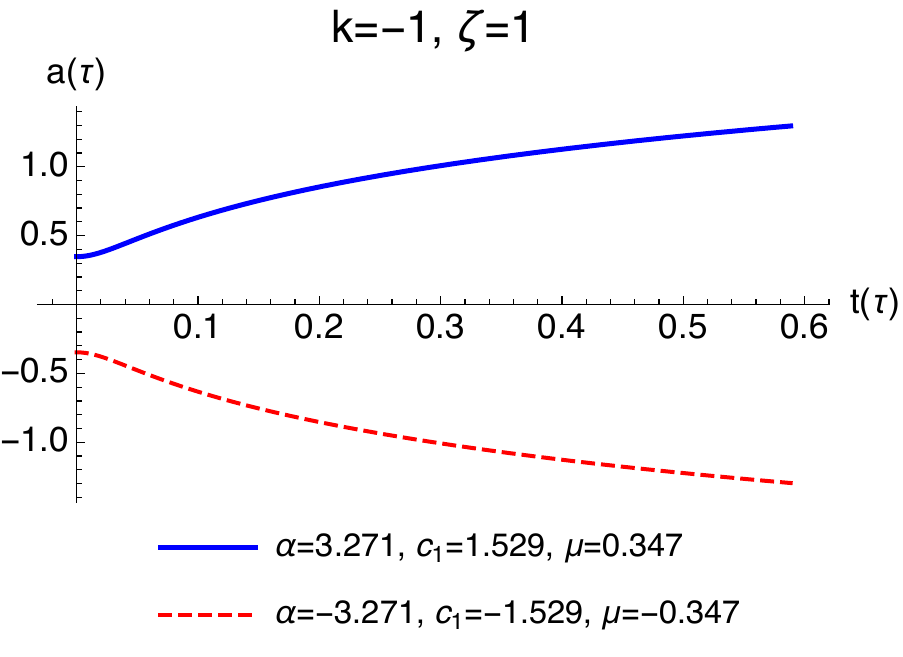}}
\caption{\label{Figura19} Rational degenerate hyperbolic solutions for superfluid closed and open universe from (\ref{eq19}), left and right, respectively.}
\end{figure}

As shown in Fig.~\ref{Figura19}, these are solutions reaching quickly plateau regions for both types of
curved spaces.

\medskip

%%%%%%%%SUBFIGURE 19%%%%%%%%%%%%%%

\begin{figure}[h!]
\centering
\subfigure[\ $a(\tau)=\mu +\frac{6.116-4\mu^3-9.813\mu^2}{2\left(-1+3\mathrm{csc}^2\left(\frac{\sqrt{6}}{2}\,\tau\right)\right)+2\mu^2+3.271\mu}$.]{
\includegraphics[scale=0.575]{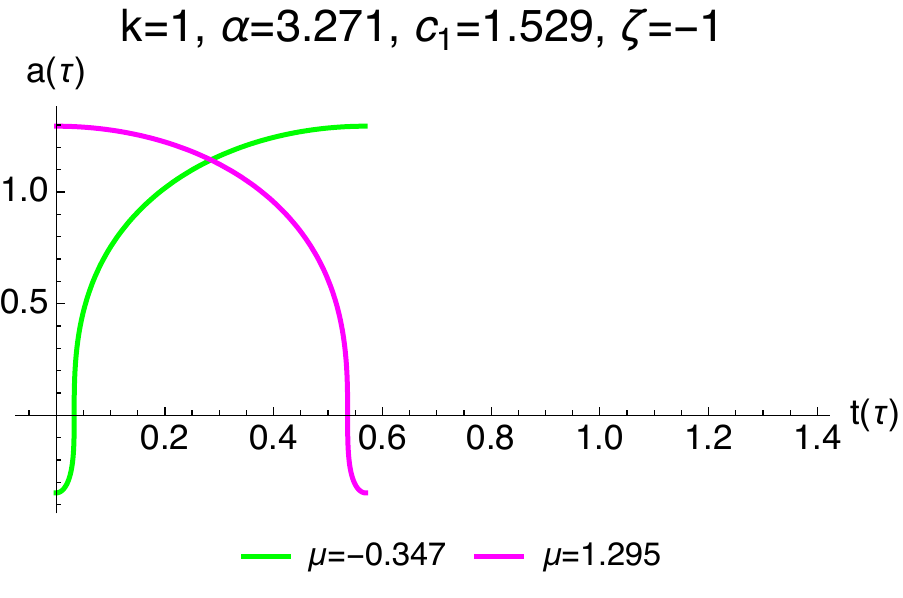}}
\subfigure[\ $a(\tau)=\mu +\frac{7+4\mu^3-15\mu^2}{2\left(-1+3\mathrm{csc}^2\left(\frac{\sqrt{6}}{2}\sqrt[3]{2}\,\tau\right)\right)-4\mu^2+5\mu}$.]{
\includegraphics[scale=0.575]{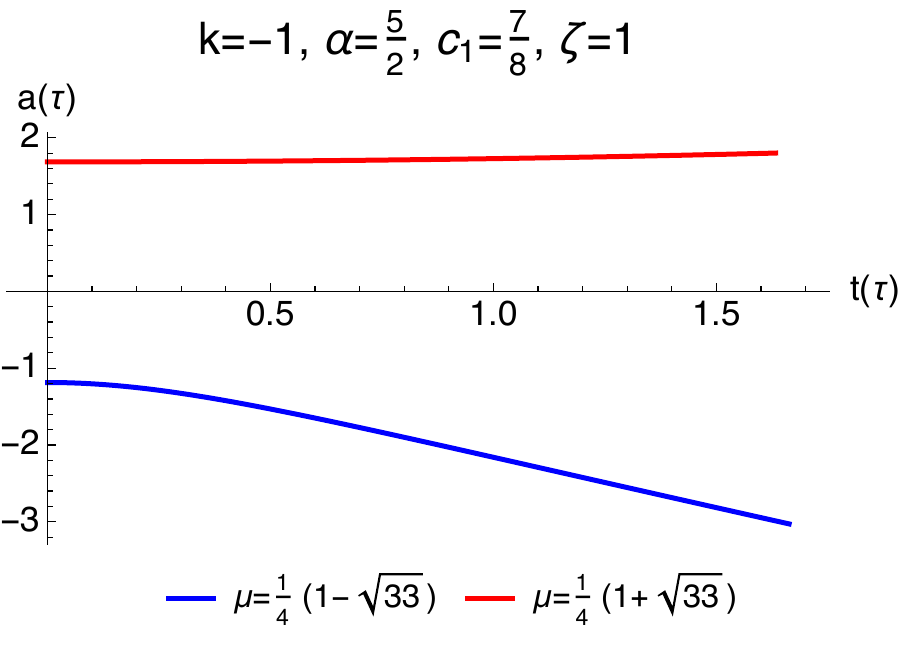}}
\caption{\label{Figura21} Rational degenerate trigonometric solutions for superfluid closed and open universe from (\ref{eq21}).}
\end{figure}

{\bf ii)}   Let $g_2=12{\tilde {e}}^2>0$ and $g_3 =8{\tilde {e}}^3>0$;  the Weierstrass $\wp$ function is  simplified to
$\wp(\tau;12{\tilde {e}}^2,8{\tilde {e}}^3)=\tilde e \left[-1+3~ \mathrm{csc}^2(\sqrt{3{\tilde {e}}}~\tau)\right]$.
Thus, we obtain the rational {\it trigonometric} solutions
\begin{equation}  \label{eq21}
a(\tau)=\mu+\frac{4c_1^*-4 k  \mu^3-3\alpha \mu ^2}{ 2\sqrt[3] {g_3}\left[-1+3\mathrm{csc}^2\left(\frac{\sqrt 6}{2}\sqrt[6]{g_3}(\tau-\tau_0)\right)\right]+2 k{\mu}^2+\alpha \mu}~, %\quad k \ne 0, \quad \Xi = 0,
\quad g_2>0, \quad g_3>0.
\end{equation}
As shown in the plots of Fig.~\ref{Figura21}, the closed universe case can have contributions from the initially unphysical trigonometric solution.

\medskip

%%%%%%%%%SUBFIGURE 21%%%%%%%%%%%%%%

{\bf iii)} When both germs are zero then $\wp(\tau,0,0)=\frac{1}{\tau^2}$, and we have the {\it rational} solution
\begin{equation}\label{eq22}
a(\tau)=\mu+\frac{Q_a(\mu)\zeta^2}{4- \frac{Q_{aa}(\mu)}{6}\zeta^2}~,
\end{equation}
which simplifies to
\begin{equation}\label{eq23}
a(\tau)=\mu+\frac{(4c_1-4 k  \mu^3-3\alpha \mu ^2)\tau^2}{4+(2 k{\mu}^2+\alpha \mu)\tau^2}~, %\quad k \ne 0, \quad \Xi = 0
\quad g_2=g_3=0~. %\quad g_3=0.
\end{equation}
Moreover, this case leads to $c_1=\frac{\alpha^3}{16 k^2}$, and $\zeta=-\frac{\alpha^4}{32 k^3}$, %\tcb{
thus the quartic polynomial has four real zeros inferred from $Q(\mu)=\frac{1}{16 k^3}(\alpha-2 k \mu)(\alpha+2 k \mu)^3$,
%%has two real roots of multiplicity two given by  $Q(\mu)=(\mu+\alpha/2)^2(\mu-\alpha/2)^2$
so the graph of \eqref{eq23} becomes the %%{\it Witch of Agnesi}
{\it Agnesi} curve \cite{Maria}
\begin{equation}\label{eq24}
\begin{aligned}
a(\tau)&=\frac{\alpha}{2}-\frac{\alpha^3 \tau^2}{4+\alpha^2 \tau^2}=\frac{\alpha}{2}\left(1-\alpha \tau^2\frac{d}{d\tau}\mathrm{arctan}(\alpha\tau/2)\right), \quad k=1, %\quad \Xi = 0
\quad g_2=g_3=0~,\\ %\quad g_3=0.\\
a(\tau)&=-\frac{\alpha}{2}-\frac{\alpha^3 \tau^2}{4-\alpha^2 \tau^2}=-\frac{\alpha}{2}\left(1+\alpha \tau^2\frac{d}{d\tau}\mathrm{arctanh}(\alpha\tau/2)\right), \quad k=-1,
\quad g_2=g_3=0~.\\
\end{aligned}
\end{equation}

\begin{figure}[h!]
\centering
\subfigure[\ $a(\tau)=\pm\frac{1}{2}\left(1\pm \tau^2\frac{d}{d\tau}\mathrm{arctan}(\frac{\tau}{2})\right)$.]{
\includegraphics[scale=0.575]{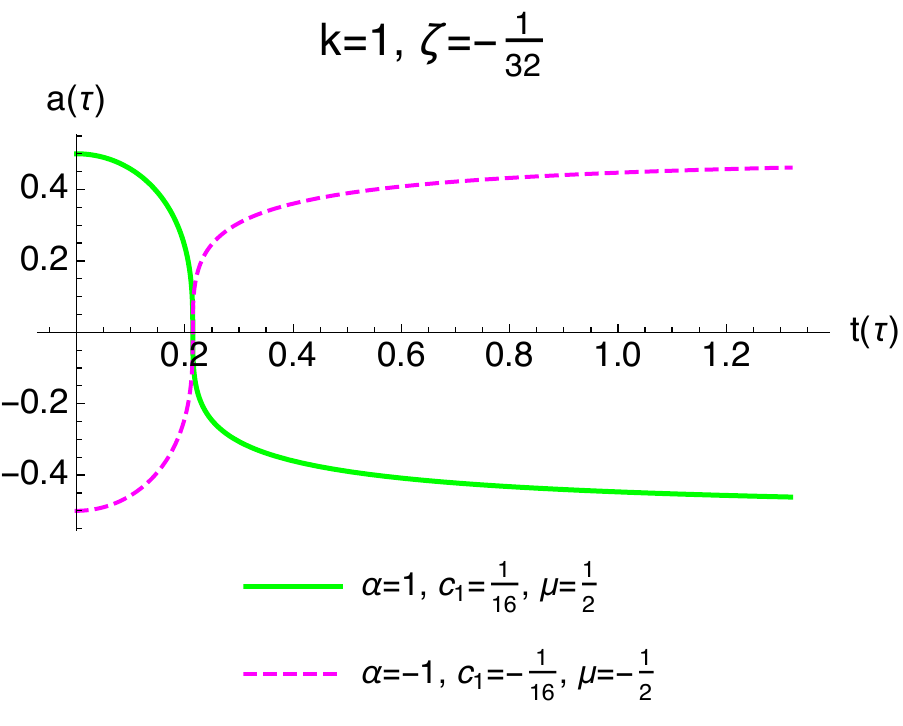}}
\subfigure[\ $a(\tau)=\pm\frac{1}{2}\left(1+\tau^2\frac{d}{d\tau}\mathrm{arctanh}(\frac{\tau}{2})\right)$.]{
\includegraphics[scale=0.575]{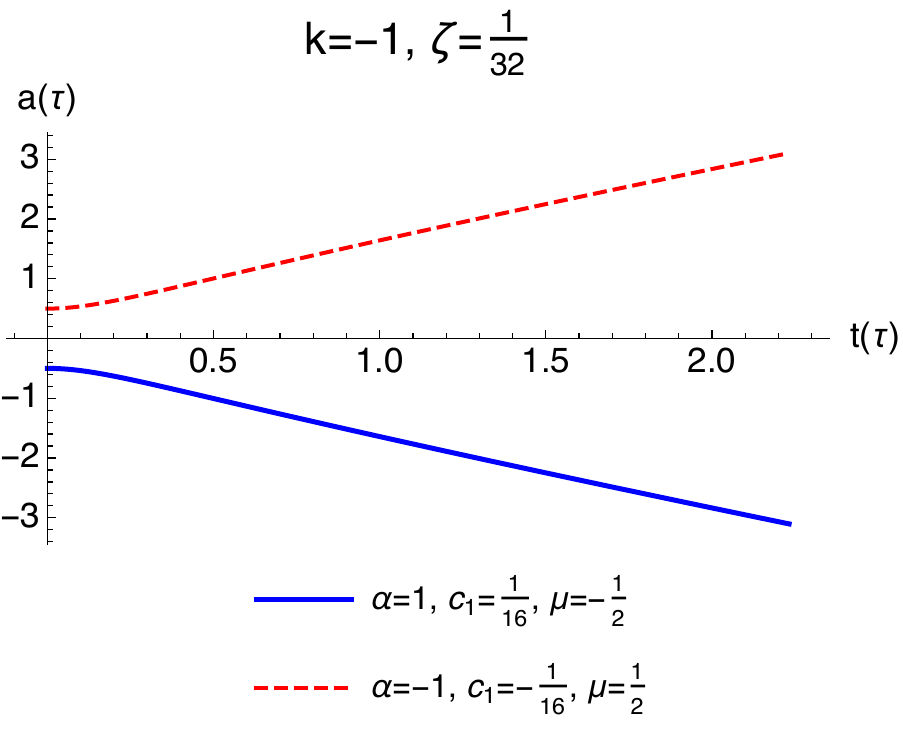}}
\caption{\label{Figura24} Degenerate Agnesi solutions for superfluid closed and open universe from (\ref{eq24}), left and right, respectively.}
\end{figure}

The degenerate Agnesi solutions in curved spaces are presented in Fig.~\ref{Figura24}, and again one can see that it is possible that the initially negative solution can become the physical solution after some time in the case of closed universes.

%%%%%%%%%SUBFIGURE 24%%%%%%%%%%%%%%

\medskip
For all cases, the comoving time in parametric form is obtained from  $dt =a^2 d\tau$ which gives
then
\begin{equation}\label{eq25}
t(\tau)=\int_{\tau_0}^{\tau} a^2(\xi) d\xi~.
\end{equation}

\subsection{Flat Space Cases, $k=0$, $\Xi\neq 0$}
In this case, \eqref{eq6} simplifies to the  elliptic equation
\begin{equation} \label{eq26}
{a_\tau}^2=-\alpha a^3 +4c_1 a-2 \zeta
\end{equation}
which can be reduced to the Weierstrass elliptic equation \eqref{eq16} by the scale transformation
$a(\tau)=-\frac{4}{\alpha}\wp(\tau-\tau_0;g_2,g_3)$, where
\begin{equation}\label{eq27a}
g_2=c_1\alpha~, \quad g_3=\frac 1 8  \alpha^2 \zeta~, \quad  \Xi=(c_1 \alpha)^3-\frac{27}{64}(\alpha^2 \zeta)^2
\end{equation}
and so, the parametric {\it Weierstrass} solutions
for $k=0$ are
\begin{equation}\label{eq28}
\begin{aligned}
a(\tau)&=-\frac{4}{\alpha}\wp\left(\tau-\tau_0;c_1\alpha,\frac 1 8  \alpha^2 \zeta\right),
\quad \alpha \ne 0, \quad \zeta \ne 0\\
t(\tau) &=\int_{\tau_0}^\tau a^2(\xi) d\xi=\frac{4}{3\alpha^2}\bigg[c_1\alpha(\tau-\tau_0)+\hat {c}_1+
2\wp'\left(\tau;c_1\alpha,\frac 1 8  \alpha^2 \zeta\right)\bigg],  \quad \tau_0 \ne0~,\end{aligned}
\end{equation}
where $\hat c_1=2\wp'\left(\tau_0;c_1\alpha,\frac 1 8  \alpha^2 \zeta\right)$.
\medskip
We will consider only the simpler equianharmonic subcase $c_1=0$, but still describing a superfluid FLRW (flat) universe.\\

\begin{figure}[h!]
\centering
\subfigure[\ $a(\tau)=\pm 4\wp(\tau-1;0;\frac{1}{8})$.]{
\includegraphics[scale=0.575]{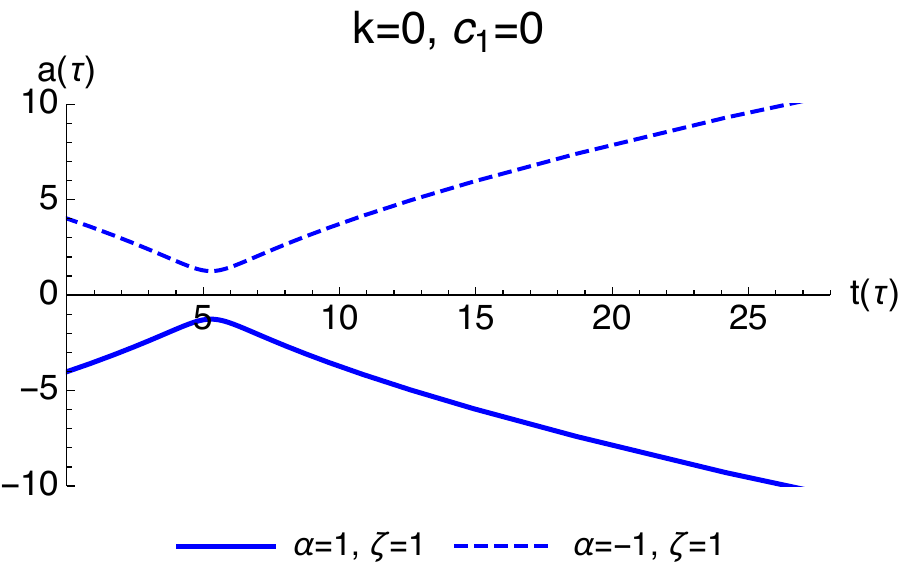}}
\subfigure[\ $a(\tau)=\pm 4\wp(\tau-1;0;-\frac{1}{8})$.]{
\includegraphics[scale=0.575]{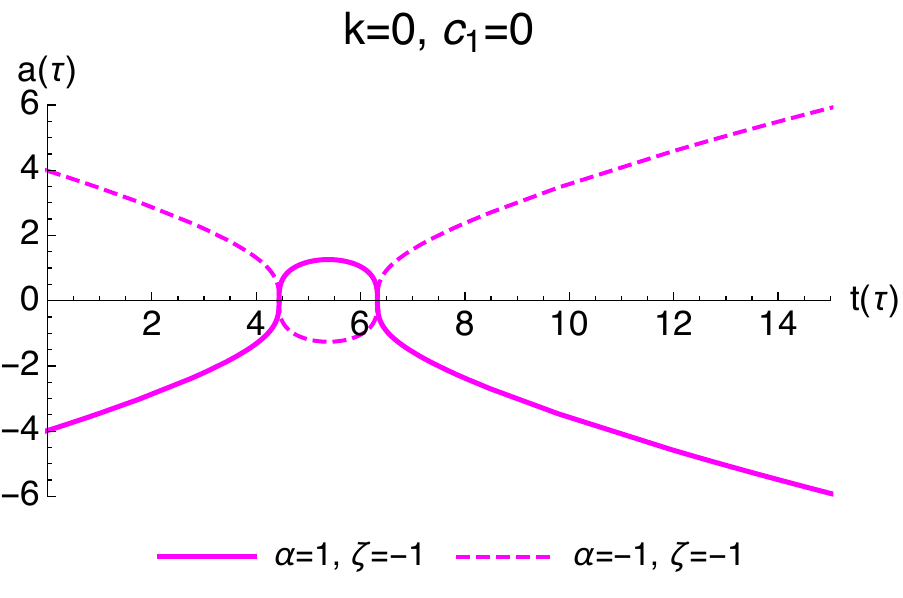}}
\caption{\label{Figura33} Reduced equianharmonic solutions for a superfluid FLRW flat space from (\ref{eq33}) for $\tau_0=1$ which avoids the singularity of $\wp'$ at the origin.}
\end{figure}

{\bf i) $c_1=0$:}
one obtains the reduced elliptic equation
\begin{equation} \label{eq32}
{a_\tau}^2=-\alpha a^3-2 \zeta~.
\end{equation}
The solutions can be obtained directly from \eqref{eq28} to give
\begin{equation}\label{eq33}
\begin{aligned}
a(\tau)&=-\frac{4}{\alpha}\wp\left(\tau-\tau_0;0,\frac 1 8  \alpha^2 \zeta\right)~,
\quad \alpha \ne 0, \quad \zeta \ne 0, \quad c_1 =0,\\
t(\tau) &=\int_{\tau_0}^\tau a^2(\xi) d\xi=\frac{8}{3\alpha^2}\bigg[\hat {c}_1+
\wp'\left(\tau;0,\frac 1 8  \alpha^2 \zeta\right)\bigg], \quad \tau_0 \ne0,\end{aligned}
\end{equation}
where $\hat c_1=\wp'\left(\tau_0;0,\frac 1 8  \alpha^2 \zeta\right)$. For this reduced {\it equainharmonic} case it is also possible to have
physical contributions from the initially negative solution as seen in Fig.~\ref{Figura33}.

%%%%%%%%%FIGURE 33%%%%%%%%%%%%%%

 \subsection{Degenerate Modular Cases, $k=0$, $\Xi=0$}
These cases are obtained for $c_1^*=\frac 3 4 \sqrt[3]{\alpha \zeta^2}$ which leads to hyperbolic or trigonometric solutions, or when $\alpha=0$ which is physical as standard FLRW only when $\zeta=0$.\\
{\bf i)}   Let $g_2=12{\hat e}^2=\frac 3 4 \alpha^{4/3}\zeta ^{2/3}>0$ and $g_3=-8{\hat e}^3=\frac 1 8 \alpha^2 \zeta<0$. These values of the germs lead to
the  {\it hyperbolic} solutions
\begin{equation}  \label{eq34}
a(\tau)=-\frac 2 \alpha \sqrt[3] {-g_3}\left[1+3\mathrm{csch}^2\left(\frac{\sqrt 6}{2}\sqrt[6]{-g_3}(\tau-\tau_0)\right)\right]~,
%\quad k = 0, \quad \Xi = 0,
\quad g_2>0, \quad g_3<0~.
\end{equation}

{\bf ii)}   Let $g_2=12{\tilde {e}}^2=\frac 3 4 \alpha^{4/3}\zeta ^{2/3}>0$ and $g_3 =8{\tilde {e}}^3=\frac 1 8 \alpha^2 \zeta>0$. For these germs, one obtains  the {\it trigonometric} solutions
\begin{equation}  \label{eq35}
a(\tau)= -\frac 2 \alpha \sqrt[3] {g_3}\left[-1+3\mathrm{csc}^2\left(\frac{\sqrt 6}{2}\sqrt[6]{g_3}(\tau-\tau_0)\right)\right]~,
%\quad k =0, \quad \Xi = 0,
\quad g_2>0, \quad g_3>0.
\end{equation}

\medskip

{\bf iii) $\alpha=0$} and {\bf $\zeta=0$:} then \eqref{eq26} becomes $a_\tau^2= 4c_1a$
with parametric {\it polynomial} solutions
\begin{equation}\label{eq39}
\begin{aligned}
a(\tau)&=\pm 2\sqrt{c_1}(\tau-\tau_0)~, %\quad k=0, \quad \Xi \ne0,
\quad \alpha=  0, \quad \zeta = 0  \\
t(\tau)&= \frac 43 c_1 \tau (\tau^2-3 \tau \tau_0+3 {\tau_0}^2)~.\\
\end{aligned}
\end{equation}

\begin{figure}[htpb]
\centering
%%\subfigure[\ The lemniscatic solution $a(\tau)=2\mathrm{cn}^2\left(\tau\left|\frac{\sqrt 2}{2}\right.\right)$ from (\ref{eq31}).]{
%%\includegraphics[scale=0.575]{fig31.pdf}}
\subfigure[\ Degenerate hyperbolic solutions in flat space $a(\tau)=\pm\left(1+3\mathrm{csch}^2\left(\frac{\sqrt{3}}{\sqrt[6]{2^5}}\tau\right)\right)$ from (\ref{eq34}).]{
\includegraphics[scale=0.575]{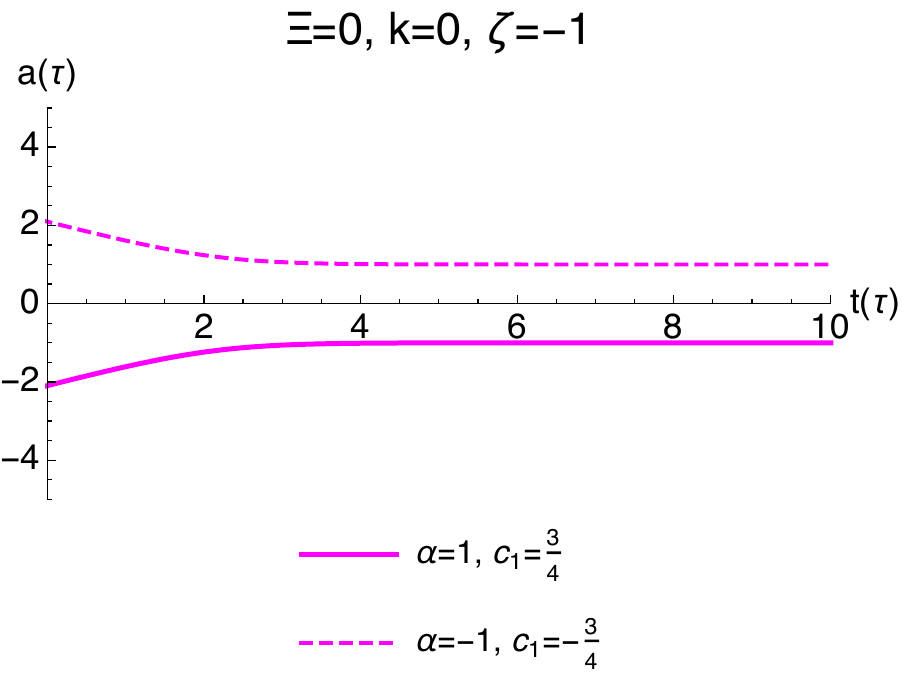}}
\subfigure[\ Degenerate trigonometric solutions in flat space $a(\tau)=\pm\left(-1+3\mathrm{csc}^2\left(\frac{\sqrt{3}}{\sqrt[6]{2^5}}\tau\right)\right)$ from (\ref{eq35}).]{
\includegraphics[scale=0.575]{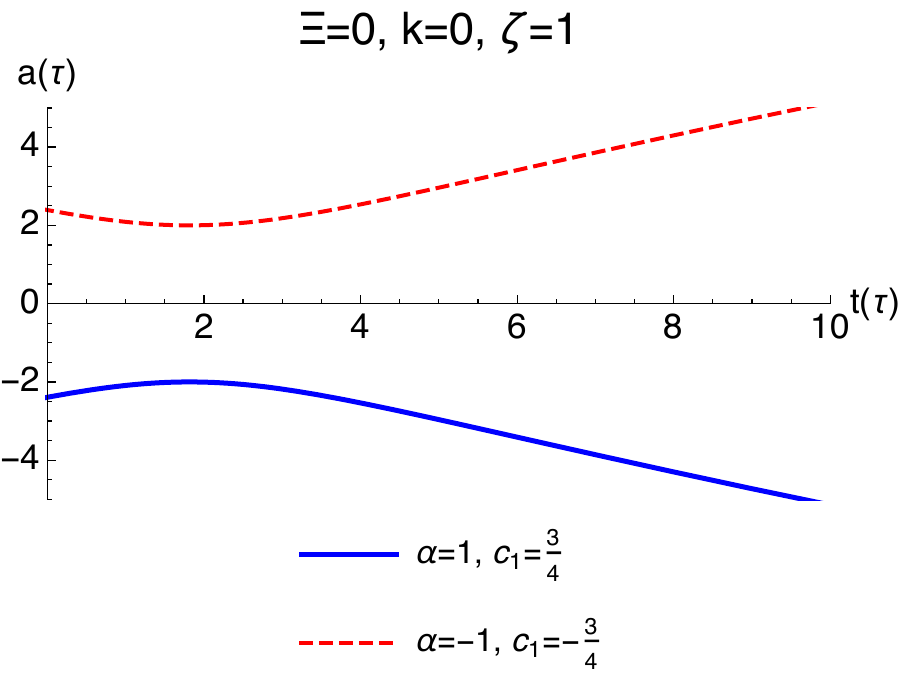}}
\subfigure[\ Degenerate polynomial solutions in flat space $a(\tau)=\pm 2\tau$ from (\ref{eq39}).]{
\includegraphics[scale=0.575]{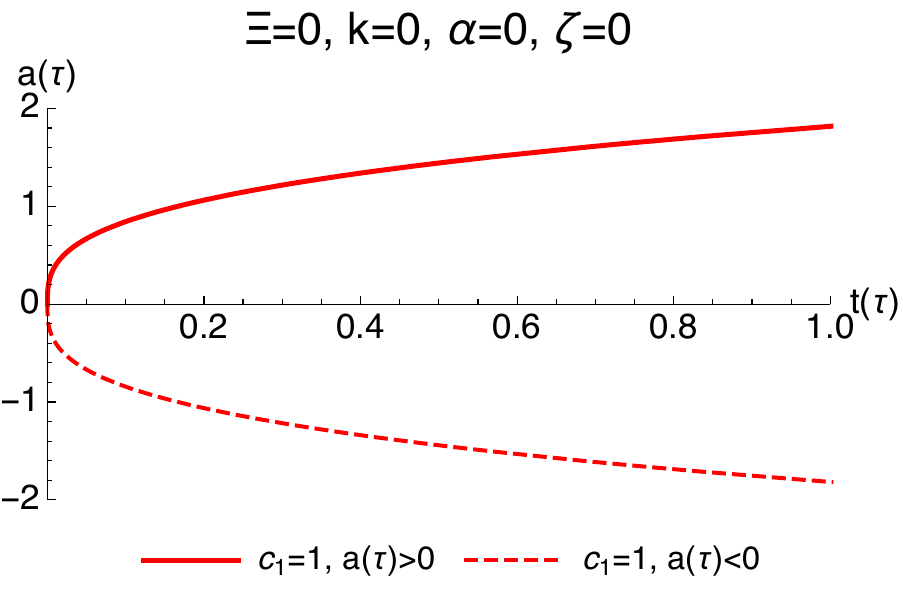}}\\
\caption{\label{Figura31343539} Modular degenerate solutions of hyperbolic, trigonometric, and polynomial type corresponding to the superfluid FLRW flat space.}
%%from (\ref{eq31}), (\ref{eq34}), (\ref{eq35}), (\ref{eq39}).}
\end{figure}

Examples of the degenerate hyperbolic, trigonometric, and the last class of polynomial solutions in flat space are illustrated in Fig.~\ref{Figura31343539} without showing contributions from the initially negative solutions for the employed parameters.

%%%%%%FIGURES 31 34,35,39

\subsection{Estimating the growth of the solutions near the origin}
In the Sundman time variable, the above solutions do not show quasi-de Sitter behavior near the origin.
Indeed, since the majority of the solutions are of the type
 \begin{equation}\label{est1}
 a(\tau) \sim \frac{p_3(\mu)}{4\wp(\tau) +p_2(\mu)}
 \end{equation}
 where $p_3$ and $p_2$ are polynomials of order three and two in $\mu$, we
 use the well-known series expansion
 \begin{equation}\label{est2}
 \wp(\tau)=\tau^{-2}+\sum_{k=2}^{\infty}c_k \tau^{2k-2}~,
 \end{equation}
 where $c_2=g_2/20$, $c_3=g_3/28$, and $c_k=\frac{3}{(2k+1)(k-3)}\sum_{m=2}^{k-2}c_mc_{k-m}$.
 Close to the origin, we can approximate $\wp \sim 1/\tau^2$, and (\ref{est1}) becomes
 \begin{equation}\label{est3}
 a(\tau)\sim \frac{p_3(\mu)\tau^2}{4 +p_2(\mu)\tau^2}~,
 \end{equation}
 which is only a slowly increasing or decreasing function depending on the values taken by $\mu$.

 \medskip

 There are also cases, such as the reduced equianharmonic one, where
 \begin{equation}\label{est2}
 a(\tau)\sim \wp(\tau;0,g_3)~.
 \end{equation}
 The series expansion for the equianharmonic case is
 \begin{equation}\label{est2}
 \wp(\tau;0,g_3)=\frac{1}{\tau^2}+c_{e,3}\tau^4+c_{e,6}\tau^{10}+ ...
 \end{equation}
 so unfolding from the origin, there is a $\tau^{-2}$ decrease up to a minimum of $a$ that can be calculated with some approximation
 from $\wp'=0$ by derivation of the three terms above followed by a multiple-power increase.

\section{The conformal time approach}\label{sec4}

The rational solutions in terms of Weierstrass elliptic functions, and degenerate cases that we introduced above owe their existence
to the usage of the Sundman time variable. They have the peculiarity that they belong only
to the case $\chi=3/2$ and besides the majority of them occur for the integration constant $c_1\neq 0$. This is due to the fact that the integrating factor method works only for the three halves value of the parameter $\chi$. It is natural to ask what happens for other values of $\chi$.
In this section we deal with this issue, which, also nontrivial, is much closer to the vast existing cosmological literature in the sense that the Rayleigh-Plesset modified gravity equation \eqref{eq4} written for a general $\chi$ parameter as
\begin{equation}\label{comuv1}
a\ddot{a}+\chi\dot{a}^2=\zeta a^{-4}-\alpha a^{-1}-\chi k~,
\end{equation}
can be approached by the usual change of variable to conformal time $\eta$, defined by $dt=a d\eta$, in which it takes the form
\begin{equation}\label{conf1}
a a_{\eta\eta}+(\chi-1){a_\eta}^2=\zeta a^{-2}-\alpha a-\chi k a^2~.
\end{equation}
A further change of dependent variable $a=b^{1/\chi}$, $\chi\neq 0$, leads to the equations
 \begin{equation}\label{b1}
 b_{\eta\eta}=\chi \bigg[\zeta b^{1-\frac{4}{\chi}}-\alpha b^{1-\frac{1}{\chi}}-\chi k b\bigg]~.
 \end{equation}
%% and so the embarrassing square derivative
For $\chi=-1$, which is the vacuum case, we obtain the equations
 \begin{equation}\label{bchi-1}
 b_{\eta\eta}\pm b-\alpha b^{2}+\zeta b^{5}=0~, \qquad b_{\eta\eta}-\alpha b^{2}+\zeta b^{5}=0~,
\end{equation}
for $k=\pm 1$ and $k=0$, respectively.
%\textcolor{red}{These are equations with negative quadratic nonlinearity and a fifth power MEB nonlinearity that can be treated perturbatively--here we do numerically .} \textcolor{blue}{
Once the  solutions  of  (\ref{bchi-1}) are found the scale factor in the conformal time is obtained through $a=\frac 1 b$.

\medskip

The conservative form of (\ref{b1}) can be easily obtained by multiplying by $b_\eta$ and after a quadrature leads to
  \begin{equation}\label{bchi-conserv}
  {b_{\eta}}^{2}+V(b)=E
 \end{equation}
 where the integration constant $E$, playing the role of the total energy, can be set to zero in the cosmological context, while the potential energy
 is given by
 \begin{equation}\label{Vb}
 V(b)=\chi^2\bigg[kb^2+\frac{2\alpha}{2\chi-1}b^{\frac{2\chi-1}{\chi}}-\frac{\zeta}{\chi-2}b^{\frac{2(\chi-2)}{\chi}}\bigg]~.
 \end{equation}

\begin{figure}[htpb]
\centering
%%\subfigure[\ The lemniscatic solution $a(\tau)=2\mathrm{cn}^2\left(\tau\left|\frac{\sqrt 2}{2}\right.\right)$ from (\ref{eq31}).]{
%%\includegraphics[scale=0.575]{fig31.eps}}
\subfigure[\ ]{
\includegraphics[scale=0.830]{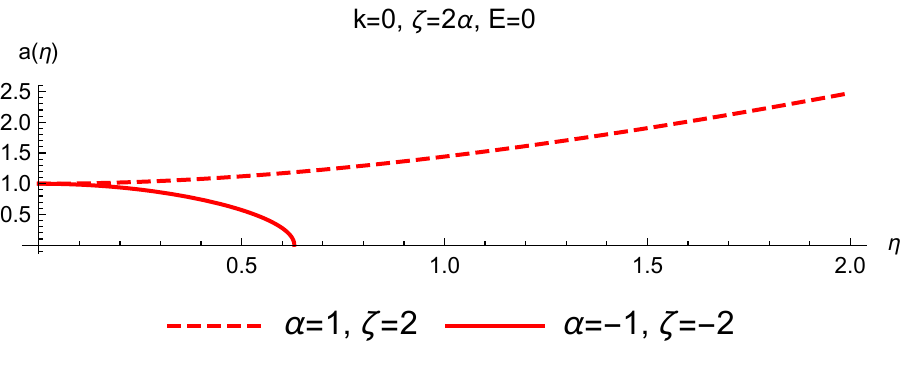}}
\subfigure[\ ]{
\includegraphics[scale=0.830]{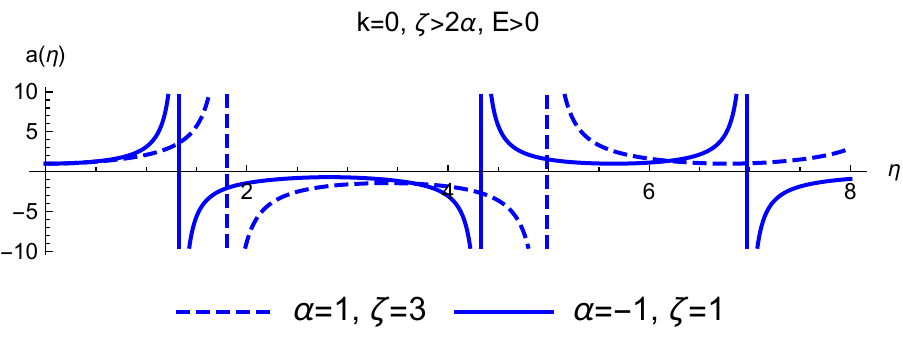}}
\subfigure[\ ]{
\includegraphics[scale=0.830]{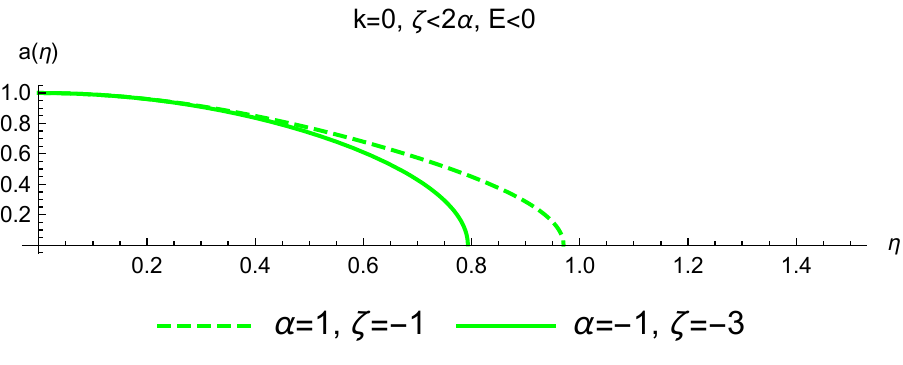}}\\
\caption{\label{Figura10} Scale factors $a(\eta)$ obtained by numerical solutions of Eq.~(\ref{bchi-1}) for the vacuum case $\chi=-1$.}
%%from (\ref{eq31}), (\ref{eq34}), (\ref{eq35}), (\ref{eq39}).}
\end{figure}

For  $\chi=-1$, we obtain
\begin{equation}
{b_\eta}^2+\frac{\zeta}{3}   b^6-\frac{2}{3} \alpha  b^3+k b^2=E~,
\end{equation}
and using the initial conditions $b(0)=1$,  $b_{\eta}(0)=0$, the energy is
\begin{equation}
E=k+\frac{\zeta-2\alpha}{3}~.
\end{equation}
%\textcolor{blue}{here
The latter formula shows that one can choose the parameters such as to have a zero total energy; for example, for
flat space, this happens for $\zeta=2 \alpha$. Plots of the case $\chi=-1$ for the three cases of positive, zero, and negative energy are presented in Fig.~\ref{Figura10} for the flat space.
Notice that in the case $\zeta>2\alpha$, i.e. positive energy, there are future singularities.
Similar plots have been obtained for closed and open universes with future singularities only in the positive energy case.

%%%%%%FIGURE-1 NEW SECTION

\begin{figure}[htpb]
\centering
%%\subfigure[\ The lemniscatic solution $a(\tau)=2\mathrm{cn}^2\left(\tau\left|\frac{\sqrt 2}{2}\right.\right)$ from (\ref{eq31}).]{
%%\includegraphics[scale=0.575]{fig31.eps}}
\subfigure[\ ]{
\includegraphics[scale=0.830]{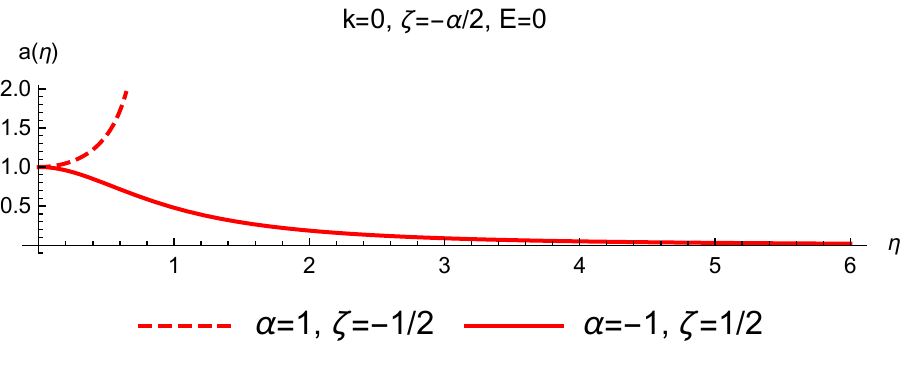}}
\subfigure[\ ]{
\includegraphics[scale=0.830]{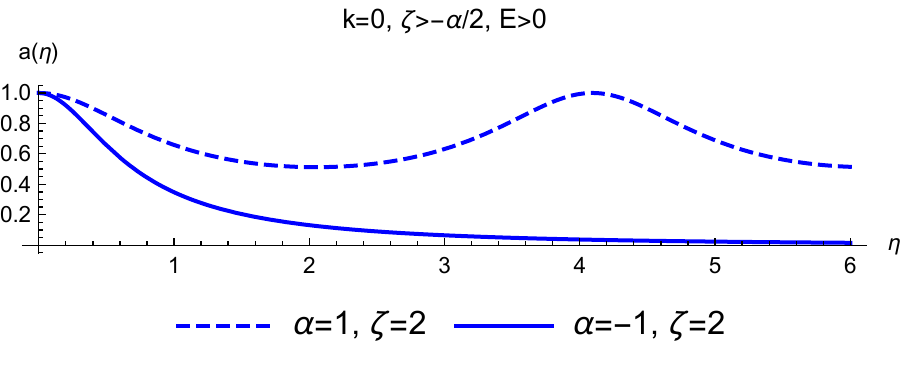}}
\subfigure[\ ]{
\includegraphics[scale=0.830]{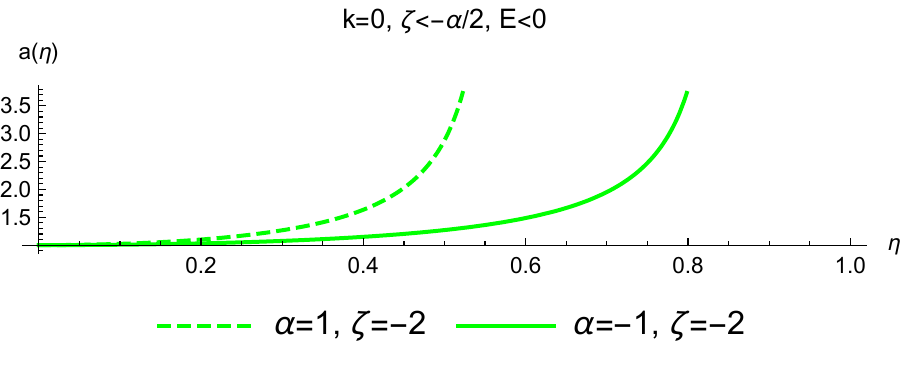}}\\
\caption{\label{Figura11} Scale factors $a(\eta)$ for the coasting case $\chi=3/2$ obtained from the numerical solutions of the $b$ equation (\ref{beqc}).}
%%from (\ref{eq31}), (\ref{eq34}), (\ref{eq35}), (\ref{eq39}).}
\end{figure}

Finally, let us briefly present numerical solutions of the scaling factors for the coasting case $\chi=3/2$ in conformal time. The $b$ equation reads
\begin{equation}\label{beqc}
b_{\eta\eta}=\frac{3}{2}\left(\zeta b^{-5/3}-\alpha b^{1/3}\right)~
\end{equation}
and the energy is
\begin{equation}\label{ecoast}
E=\frac{9}{4}\left(\alpha+2\zeta\right)~.
\end{equation}
for initial conditions as in the previous case.
In Figure~\ref{Figura11}, we display the numerically obtained scale factors for the three energy cases. Interestingly, we do not find future singularities in
the coasting case for positive energy, as in the vacuum case.

%%%%%%FIGURE-2 NEW SECTION

\section{Discussion and Conclusion}\label{sec5}
We have introduced and analyzed in some detail a nonlinear differential equation for a modified FLRW cosmology for zero cosmological constant, which is based on an analogy with a Rayleigh-Plesset equation encountered in the area of multielectron bubbles in superfluid He-4, a well studied laboratory phenomenon. At the speculative level, one can assume the existence of equivalents of such bubbles even at astrophysical scales, as a kind of microcavities hovering in the galactic halos where they could be generated due to some unknown astrophysical mechanism.

\ms

In the case of coasting universe, $\chi=3/2$, for such a modified FLRW gravity equation, we have presented parametric analytical solutions which are mainly in rational forms containing elliptic Weierstrass $\wp$ functions in terms of a variable that we call Sundman time whose differential is $a^{-2}$ times the differential of the comoving time.
We recall that for standard FLRW cosmology, some results in the comoving time involving the Weierstrass $\wp$ function go back to Lemaître himself \cite{Lmtr33}, while Coquereaux \cite{Coq15} obtained an equianharmonic (nonrational) $\wp$ solution in conformal time when the radiative contribution is neglected. The reduction to the standard FLRW case of our superfluid RP-modified FLRW cosmology provides the rational equianharmonic solution $a(\tau)=c_1/\wp(\tau;0,kc_1^2)$ for $\mu=0$ which exists only if $c_1\neq 0$ , see equation~(\ref{eq16bis}) and Figs~3(a) and 4(a). A similar result in conformal time can be found in the work of Steiner \cite{Steiner2007}, but with $\wp$ not strictly equianharmonic.
However, this author claims a correction of sign in the main formula of the first Biermann-Weierstrass theorem in \cite{Whi} that we do not confirm.

 \ms

For some particular values of the parameters, we have displayed periodic (cyclic) solutions specific to rational expressions of Weierstrass functions. The interesting feature of physical contributions from an initially nonphysical (negative) solution has been revealed, mainly but not only, for the closed universe cases. In the case in which some of these superfluid-modified scale factors will be taken into account for possible cosmological scenarios or for astrophysical phenomena, then the parameters should be fitted to more realistic values according to the sets of observational data. Furthermore, at the terrestrial laboratory level, some of the flat space solutions that have been obtained here can be also useful in the data analyses of superfluid bubble experiments.

\ms

Finally, using the more standard conformal time variable, we have also briefly discussed the $\chi=-1$ vacuum case for this Rayleigh-Plesset modification of FLRW equation and the same coasting case, although the solutions in conformal time could be obtained by numerical integration only.

\bigskip

\noindent {\bf Acknowledgement}\\

\noindent We wish to thank the anonymous referee for important remarks that helped us to improve the content of this paper.

\bigskip
\bigskip

\subsection*{Credit authorship contribution statement}

\noindent {\bf H.C. Rosu}: Writing - original draft, Supervision, Formal analysis.

\noindent {\bf S.C. Mancas}: Calculations, Writing - review \& editing.

\noindent {\bf C.-C. Hsieh}: Formal analysis, Validation.

\subsection*{Declaration of competing interests}

The authors declare they have no known competing financial interests or personal relationships that could
have appeared to influence the work reported in this paper.

\subsection*{Aknowledgements}

%We thank our institutions for financial support.
We wish to thank the anonymous referee for important remarks that helped us to improve the content of this paper.

%%\begin{figure}[h!]
%%  \includegraphics[width=\linewidth]{x.png}
%%  \caption{some words}
%%  \label{fig:fig1}
%%\end{figure}

\end{document}